\documentclass{nature}

\usepackage{graphicx}
\makeatletter
\let\saved@includegraphics\includegraphics
\AtBeginDocument{\let\includegraphics\saved@includegraphics}
\renewenvironment*{figure}{\@float{figure}}{\end@float}
\makeatother

\title{A Hexagon in Saturn's Northern Stratosphere Surrounding the Emerging Summertime Polar Vortex}

\author{L.N. Fletcher*$^{1}$, 
G.S. Orton$^2$, 
J.A. Sinclair$^2$, 
S. Guerlet$^{3}$, 
P.L. Read$^4$, 
A. Antu\~{n}ano$^1$,
R.K. Achterberg$^5$, 
F.M. Flasar$^6$,
P.G.J. Irwin$^4$ 
G.L. Bjoraker$^6$,
J. Hurley$^7$,
B.E. Hesman$^8$,
M. Segura$^6$,
N. Gorius$^9$,
A. Mamoutkine$^6$,
\& S.B. Calcutt$^4$}

\newcommand{\pderiv}[2]{\frac{\partial #1}{\partial #2}}

\usepackage{lineno} 
\usepackage[final]{changes}
\usepackage[bookmarks = true, bookmarksnumbered = true, pdfpagemode =None, pdfstartview = FitH, pdfpagelayout = SinglePage, colorlinks = true, urlcolor = blue, citecolor = blue]{hyperref}

\begin{document}

\maketitle

\begin{affiliations}
 \item Department of Physics \& Astronomy, University of Leicester, University Road, Leicester, LE1 7RH, UK (*Corresponding author leigh.fletcher@le.ac.uk).
 \item Jet Propulsion Laboratory, California Institute of Technology, 4800 Oak Grove Drive, Pasadena, CA, 91109, USA.
 \item Laboratoire de Meteorologie Dynamique/IPSL, Sorbonne Universit\'{e}, \'{E}cole normale sup\'{e}rieure, PSL Research University, \'{E}cole polytechnique, CNRS, F-75005 Paris, France.
 \item Department of Physics (Atmospheric, Oceanic and Planetary Physics), University of Oxford, Parks Rd, Oxford, OX1 3PU, UK.
 \item Department of Astronomy, University of Maryland, College Park, MD 20742, USA.
 \item NASA/Goddard Space Flight Center, Greenbelt, Maryland, 20771, USA
\item STFC Rutherford Appleton Laboratory, Harwell Science and Innovation Campus, Didcot, OX11 0QX, UK.
\item Space Telescope Science Institute (STScI), 3700 San Martin Drive Baltimore, MD 21218, USA.
\item Department of Physics, The Catholic University of America, Washington, DC 20064, USA.

\end{affiliations}


\begin{abstract}
\added{Saturn's polar stratosphere exhibits the seasonal growth and dissipation of broad, warm, vortices poleward of $\sim75^\circ$ latitude, which are strongest in the summer and absent in winter.}  The longevity of the exploration of the Saturn system by Cassini allows the use of infrared spectroscopy to trace the formation of the North Polar Stratospheric Vortex (NPSV), a region of enhanced temperatures and elevated hydrocarbon abundances at millibar pressures.  We constrain the timescales of stratospheric vortex formation and dissipation in both hemispheres.  Although the NPSV formed during late northern spring, by the end of Cassini's reconnaissance (shortly after northern summer solstice), it still did not display the contrasts in temperature and composition that were evident at the south pole during southern summer.  The newly-formed NPSV was bounded by a strengthening stratospheric thermal gradient near $78^\circ$N.  The emergent boundary was hexagonal, suggesting that the Rossby wave responsible for Saturn's \added{long-lived} polar hexagon - which was previously expected to be trapped in the troposphere - can \added{influence the stratospheric temperatures some 300 km above Saturn's clouds.}

\end{abstract}

%


\section*{Introduction}

Polar vortices are important features of planet-wide atmospheric circulation systems on terrestrial worlds and giant planets alike, \added{from Jupiter's turbulent polar environment\cite{17bolton}, to Uranus' seasonal polar hood\cite{15depater}, and Neptune's warm south polar vortex\cite{07orton_nep}.  Saturn's axial tilt of $26.7^\circ$ renders its seasonally-illuminated poles readily accessible from Earth, providing the paradigm for testing our understanding of the interplay between dynamics, chemistry, cloud formation and auroral processes in shaping giant planet polar environments}. Saturn's stratosphere, some 200-300 km above the clouds, exhibits the growth and dissipation of broad, warm polar vortices over seasonal timescales (Saturn's orbit spans 30 Earth years), which are strongest in the summer and absent during winter\cite{05orton, 08fletcher_poles}.  These vortices, which extend $\sim15^\circ$ latitude away from the pole, are bounded by a strong latitudinal temperature gradient and are distinct from the compact, seasonally-independent polar cyclones present in the troposphere\cite{06sanchez, 08fletcher_poles, 08dyudina, 09baines_pole, 15antunano, 17sayanagi}.  It has not been previously possible to constrain the formation timescales of Saturn's stratospheric vortices.  By the time the Keck observatory observed Saturn in February 2004 (a planetocentric solar longitude of $L_s=287^\circ$), 16 months after southern summer solstice (October 2002), the warm South Polar Stratospheric Vortex (SPSV) was already well established poleward of $70^\circ$S\cite{05orton}.  Half a Saturnian year earlier, when the NASA Infrared Telescope Facility observed Saturn's North Pole in March 1989 ($L_s=104.5^\circ$), 15 months after the northern summer solstice (December 1987), the warm North Polar Stratospheric Vortex (NPSV) could already be seen\cite{89gezari}.  The longevity of Cassini's exploration of Saturn (2004-2017) now provides the first complete dataset to trace the growth of the NPSV, as Saturn reached northern summer solstice ($L_s=90^\circ$) in May 2017. 

Previous work used 7-16 $\mu$m infrared spectroscopy \added{obtained by the Cassini spacecraft}\cite{15fletcher_poles} to trace the dissipation of the SPSV from northern winter (July 2004, $L_s=293^\circ$) to northern spring (June 2014, $L_s=56^\circ$).  However, the seasonal evolution in the northern hemisphere was less dramatic: although the north had been warming, as expected from radiative \added{and dynamical} climate models \cite{14guerlet, 12friedson, 08greathouse_agu}, there was no sign of the sharp latitudinal thermal gradients (and hence vertical windshear) that had been associated with the SPSV at the beginning of Cassini's mission.  This suggested\cite{15fletcher_poles} that the NPSV might begin to form during the remaining years of the Cassini mission (2014-2017, $L_s=56-93^\circ$), during late spring or early summer.

Theoretical expectations for the warming and cooling of Saturn's stratosphere are hampered by an incomplete knowledge of both the distribution of opacity sources contributing to the radiative balance\cite{85bezard, 12friedson, 14guerlet}; and of contributions from dynamical upwelling and subsidence\cite{90conrath, 12friedson}.  In particular, the primary stratospheric coolants (ethane and acetylene) are spatially and temporally variable due to both photochemistry and atmospheric circulation\cite{05moses_sat, 07moses, 09guerlet, 10guerlet, 13sinclair, 15sylvestre, 15fletcher}, and the rarity of studies  constraining the properties of polar stratospheric aerosols prevents a comprehensive assessment of their contribution to the radiative budget\cite{14guerlet, 15guerlet, 16koskinen}.  Whilst subsidence within the vortices might be expected to cause heating by adiabatic compression, the increased hydrocarbon and aerosol abundances could balance this by enhanced radiative cooling, rendering this coupled radiative-dynamical-chemical balance extremely complex.  

Here we provide observational evidence for the development of localised thermal and compositional gradients to understand the formation timescales of the NPSV, and present the final status of Saturn's polar stratosphere at the end of the Cassini mission.  Furthermore, the spatial resolution and sensitivity provided by the Composite Infrared Spectrometer (CIRS)\cite{04flasar} reveal a significant surprise:  the NPSV exhibits a hexagonal boundary that mirrors the well-studied hexagonal wave in Saturn's troposphere\cite{88godfrey, 14sanchez, 16sayanagi}.  The meandering of the jet that forms the hexagon is believed to be a Rossby wave\cite{90allison_hex} resulting from an instability of the eastward zonal jet near $78^\circ$N\cite{09read_epv, 10barbosa, 14sanchez, 15morales} and trapped within a waveguide formed by the zonal jets and Saturn's vertical static stability profile.  We show here that Saturn's famous hexagon is not \added{always restricted to the troposphere, but can persist high in the stratosphere in the spring/summer}, creating a hexagonal structure that spans more than $\sim300$ km in height from the clouds to the stratospheric polar vortex.

\section*{Results}

\subsection{Hexagonal boundary of the NPSV.}
\label{hexagon}

This study uses nadir-sounding spectroscopy from Cassini's Composite Infrared Spectrometer (CIRS)\cite{04flasar, 17jennings}, a 10-1400 cm$^{-1}$ Fourier Transform Spectrometer with programmable spectral resolution between 0.5-15.5 cm$^{-1}$. We utilise the mid-infrared Michelson interferometer (600-1400 cm$^{-1}$), which featured two arrays of $1\times10$ HdCdTe detectors, each with an instantaneous field of view of $0.27\times0.27$ mrad.  Spectra are inverted using an optimal estimation retrieval algorithm\cite{00rodgers, 08irwin} to characterise the upper tropospheric (80-300 mbar) and stratospheric (0.5-5.0 mbar) temperatures and composition\cite{07fletcher_temp} as a function of time (see Methods).  \added{Before performing spectral inversions, we begin with an investigation of the polar morphology observed in the raw brightness temperatures.}

\begin{figure}
\centering
\makebox[\textwidth][c]{\includegraphics[width=1.3\textwidth]{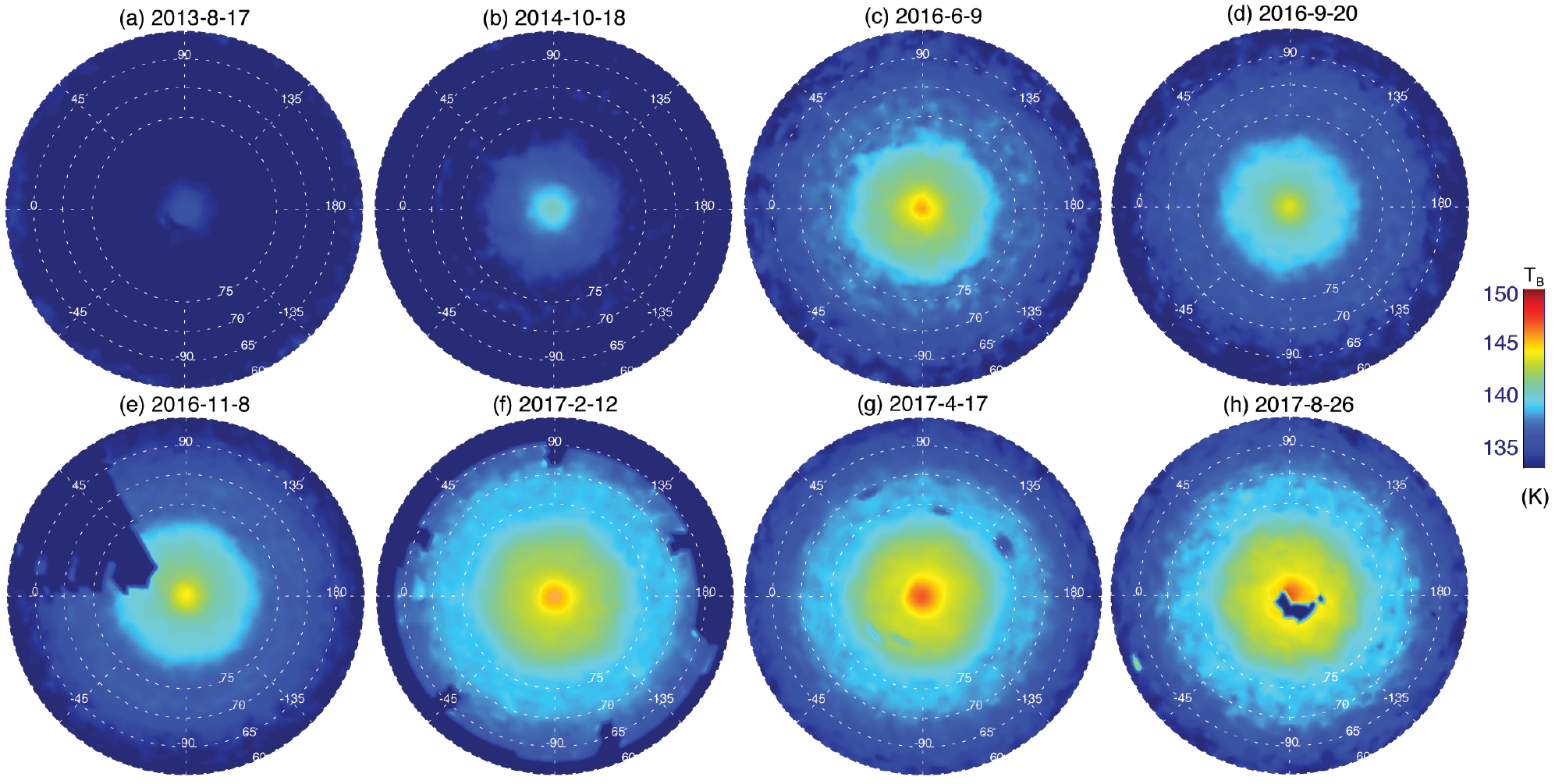}}
\caption{\textbf{Brightness temperature maps of the north polar stratosphere from 2013 to 2017.}  Brightness temperatures were averaged over the 1280-1320 cm$^{-1}$ range sensing 0.5-5.0 mbar, and each map has the same temperature scale.  The figure indicates the hexagonal boundary to the polar stratospheric hood. The panels provide maps for (a) August 17, 2013, (b) October 18, 2014, (c) June 9, 2016, (d) September 20, 2016, (e) November 8, 2016, (f) February 12, 2017, (g) April 17, 2017, (h) August 26, 2017.  Maps with spectral resolutions of 2.5 cm$^{-1}$ (REGMAPs) and 15.0 cm$^{-1}$ (FIRMAPs) were used here, spanning Cassini's solstice mission and proximal orbits.  These data were obtained from a relatively high orbital inclination to facilitate views of the north pole. Spectra have been averaged on an equal-area projection to reduce noise. Regions of dark blue represent missing data and defects - in particular, the central polar cyclone was partially obscured in the final FIRMAP of the mission acquired in August 2017 (panel h).  The hexagonal boundary near 78-80$^\circ$ latitude is clearest in the October 2014 (b), November 2016 (e) and February 2017 (f) data.  The sinusoidal variation of temperature is shown in Fig. \ref{vertices}e-f for two dates. }
\label{hexmaps}
\end{figure}

Fig. \ref{hexmaps} displays spatially resolved maps of CIRS brightness temperatures from 2013 to 2017.  These are brightness temperatures averaged over the CH$_4$ Q band emission between 1280-1320 cm$^{-1}$, and represent the kinetic temperatures over a broad range of altitudes between 0.5-5.0 mbar\cite{07fletcher_temp}.  For the purpose of these observations, the CIRS focal plane was slowly scanned from north to south as the planet rotated beneath over one or two rotations (10-20 hours).  

Four features are immediately evident in the maps.  Firstly, the brightness temperature has increased by approximately 10 K during the four years shown in this figure.  Secondly, the compact polar cyclone\cite{08fletcher_poles} is evident in the centre of each image, related to a strong negative vertical shear on the zonal jet at $87\pm1^\circ$N (see below).  Thirdly, a boundary can be seen near 78$^\circ$N in the later images, separating cooler low-latitude temperatures from warmer polar temperatures and suggesting the strengthening of the \added{temperature gradient} there.  Finally, and most surprising, the boundary appears to have a hexagonal appearance in many of these images.  Cassini demonstrated that Saturn's famous hexagonal wave\cite{88godfrey}, observed at the cloud tops ($p\sim500$ mbar) since the Voyager era, was related to hexagonal temperature contrasts reaching as high as Saturn's tropopause ($p\sim80-100$ mbar)\cite{08fletcher_poles}. However, at the conclusion of the 2004-2014 study\cite{15fletcher_poles}, the north polar stratosphere did not exhibit enough of a latitudinal temperature gradient ($\partial{T}/\partial{y}$, where $y$ is the north-south distance) to observe any contrast associated with the jet at $78^\circ$N.  This is therefore the first evidence that Saturn's hexagon extends far higher than previously thought, into the stably-stratified middle atmosphere to the 0.5-5.0 mbar level.  

There are limitations to what the CIRS observations can tell us.  An attempt to provide a spatially-resolved temperature retrieval of the stratospheric hexagon was unsuccessful, as the binned spectra did not have sufficient signal-to-noise to constrain the hexagonal shape of the temperature field as a function of altitude.  We therefore cannot tell whether the hexagon is present at 5 mbar or at 0.5 mbar, or a combination of both.  However, data presented in the next section indicates that Saturn's banded structure is clearer at 5 mbar than at lower pressures, raising the likelihood that the hexagon is only detectable in the lower stratosphere ($p>1$ mbar).  Furthermore, the low signal in the pre-2014 data means that we cannot distinguish between scenarios where the hexagon is always present in the stratosphere irrespective of season, or the hexagon is allowed to extend upwards due to the seasonally-changing temperatures as summer solstice approaches (to be discussed below).

\begin{figure}
\centering
\makebox[\textwidth][c]{\includegraphics[width=1.3\textwidth]{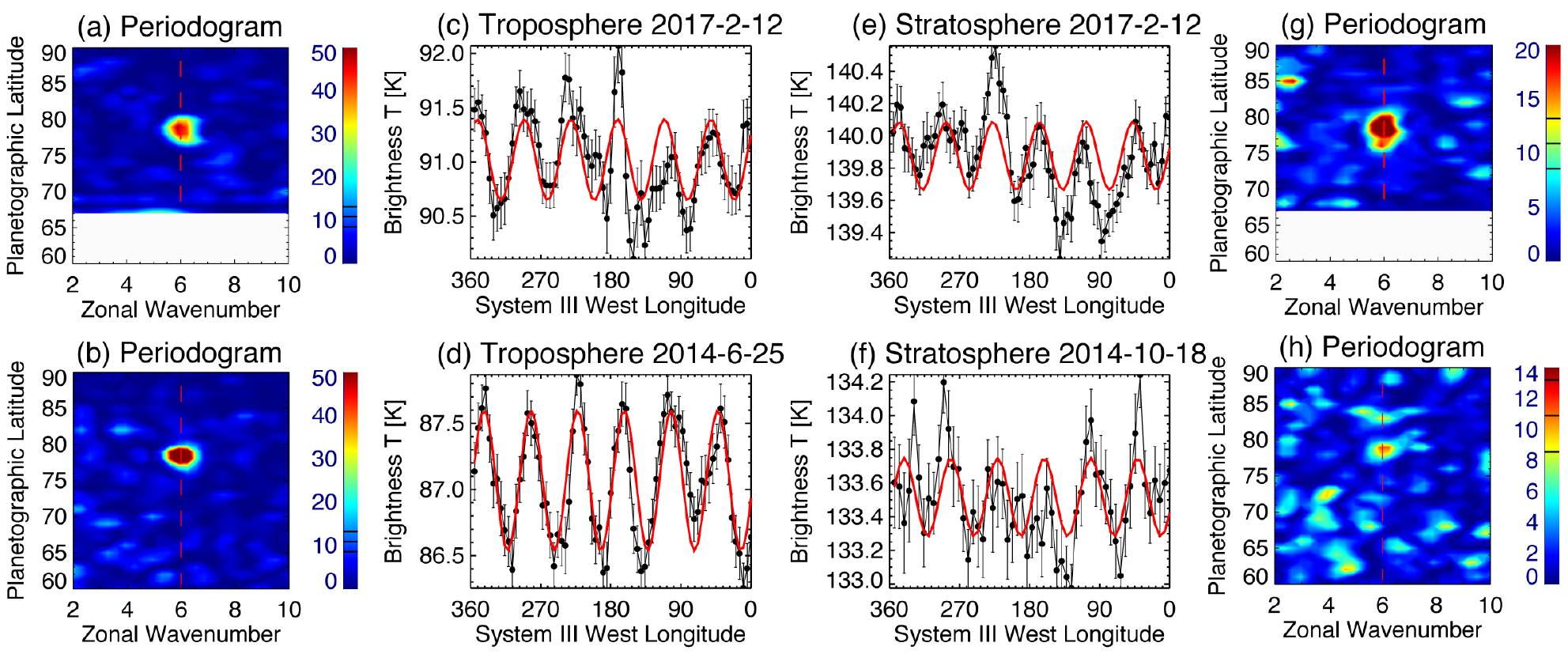}}
\caption{\textbf{Detection of the hexagon in the troposphere and stratosphere.}  Panels c-f show brightness temperatures and their uncertainties in the troposphere (c-d, averaged over 600-620 cm$^{-1}$) and stratosphere (e-f, averaged over 1280-1320 cm$^{-1}$) extracted from $77-79^\circ$N in the June 25, 2014 and February 12, 2017 REGMAP observations, and the October 18th 2014 FIRMAP observation.  As a guideline, a wavenumber-6 sinusoid has been fitted to each observation (red curve), although the stratospheric map in 2014 is extremely noisy.   Panels a, b, g and h show the corresponding Lomb Scargle periodograms\cite{08fletcher_poles}, displayed as a function of latitude, showing the spectral power contained in each longitudinally-resolved brightness temperature scan.  Black lines on the colour bars represent false-alarm probabilities (from top to bottom) of 0.001, 0.01 and 0.1.  A wavenumber-6 feature is clear in each dataset near 78$^\circ$N (vertical red dashed line), although the 2014 stratospheric hexagon is rather noisy (Fig. \ref{hexmaps}b).  Using the sinusoidal fits, we detect a westward shift in the tropospheric hexagon vertices of $8.5\pm1.1^\circ$ over 963 days (the uncertainty comes from the quality of the sinusoidal fit). Offsets between the tropospheric and stratospheric vertices were $<4^\circ$ in February 2017. }
\label{vertices}
\end{figure}

\added{Fig. \ref{vertices} displays both a periodogram analysis and an attempt to fit a simple sinusoidal shape to the hexagon brightness temperatures in 2014 and 2017.  The wavenumber-6 pattern is clearly detected in both the troposphere and stratosphere, and we find that the hexagon vertices differ in position by no more than $4^\circ$ longitude between the two altitudes, so there appears to be no \added{longitudinal} translation in the pattern with height.}  Using the tropospheric maps, we find a clockwise (westward) translation of the hexagon pattern with time, equating to $((9.9\pm1.3)\times10^{-3})^{\circ}$/day \added{(using the System III West Longitude System).  Performing the same calculation in the stratosphere yields $((7.6\pm1.9)\times10^{-3})^\circ$/day, such that the rotation rates are equivalent to within the uncertainties. Given that precise identification of the vertex longitudes is limited by the low spatial resolution and noise of the thermal maps, this drift rate is likely consistent with} the $((12.8\pm1.3)\times10^{-3})^{\circ}$/day estimate\cite{14sanchez} from tracking the hexagon at the cloud tops between 2008 and 2014.  The hexagon is therefore present \added{in a least three altitude ranges} separated by three decades of pressure, from the deep 2-3 bar troposphere\cite{09baines_pole} to the tropopause\cite{08fletcher_poles} and into the stratosphere; a vertical distance of $\sim300$ km.  \added{Finally, we note that the tropospheric hexagon vertices do not move with the $\sim120$ ms$^{-1}$ eastward cloud-top velocity of the jet at $78^\circ$N, so despite the strong windshear from the troposphere to the stratosphere (next section), the presence of a coherent wave in both altitude domains, with a coupled and very slow rotation, is plausible irrespective of the winds.  The long-term temperature changes that allowed the hexagon to be visible in the stratosphere after 2014 will be explored in the next section.  }

\subsection{Long-term temperature trends.}

\begin{figure}
\centering
\makebox[\textwidth][c]{\includegraphics[width=1.3\textwidth]{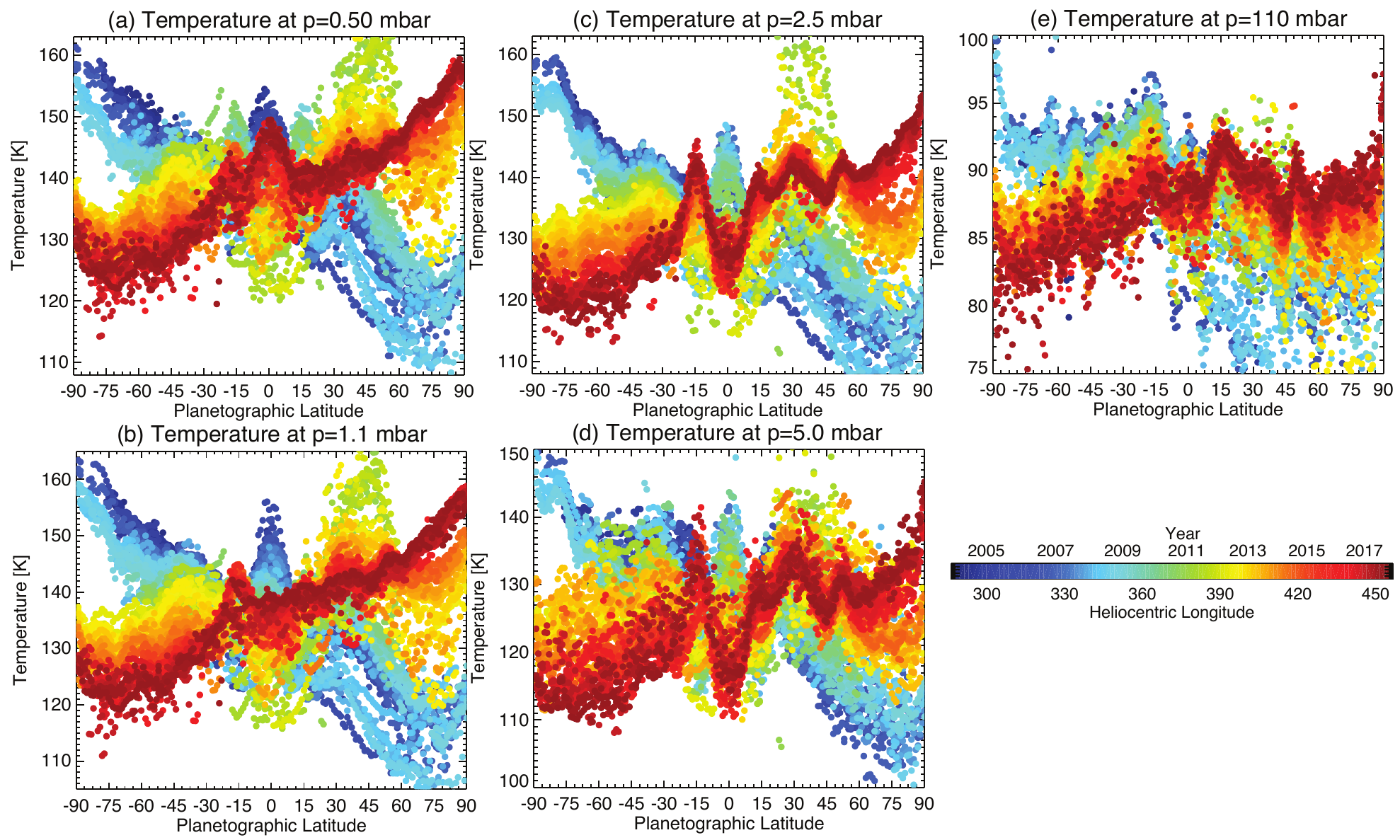}}
\caption{\textbf{Retrieved temperatures in the upper troposphere and stratosphere from the entire Cassini mission.}  Temperatures are displayed at five pressure levels:  (a) 0.5 mbar, (b) 1.1 mbar, (c) 2.5 mbar, (d) 5.0 mbar, and (e) 110 mbar. Colours refer to dates given by the legend.  Saturn's equatorial oscillation is observed at low latitudes\cite{17fletcher_qpo}, the effects of the Great Northern Storm are visible at northern mid-latitudes\cite{12fletcher}, and the reversal of the stratospheric temperature asymmetry is most evident at the poles (this work).}
\label{Tpoints}
\end{figure}

\begin{figure}
\makebox[\textwidth][c]{\includegraphics[width=1.3\textwidth]{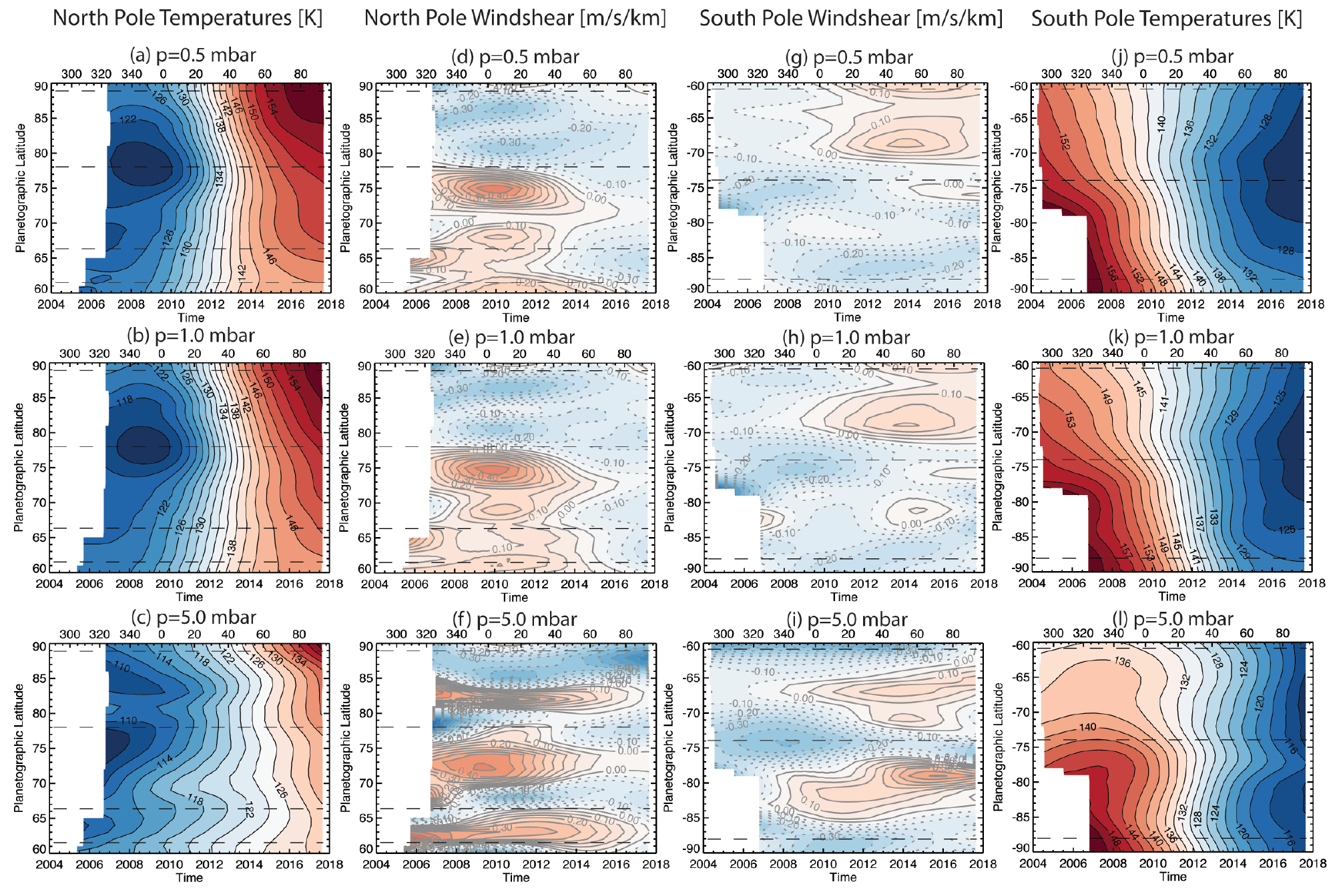}}
\caption{\textbf{North and South polar stratospheric temperature gradients as a function of time throughout the whole Cassini mission, 2004-2017.}  We display north polar temperatures (panels a-c), north polar windshears (panels d-f), south polar windshears (panels g-i) and south polar temperatures (panels j-l) at three different pressure levels (0.5, 1.0 and 5.0 mbar).  These were derived from averages of low-resolution CIRS spectra on a monthly temporal grid, and interpolated using tensioned splines\cite{07teanby_splines} to reconstruct a smoothed temperature field.  Horizontal dashed lines signify the peak of eastward zonal jets in the troposphere\cite{15antunano}.  The data are displayed as a function of time (years), but a second horizontal axis provides the planetocentric solar longitude ($L_s$) in degrees.}
\label{stratos_temp}
\end{figure}

Fig. \ref{Tpoints} displays the temperatures retrieved for each latitude \added{(averaging spectra over longitude, see Methods)} for a selection of stratospheric pressures as a function of time.  These are interpolated in Fig. \ref{stratos_temp} to display the reconstructed temperatures for the North Pole and the South Pole for three altitudes (0.5, 1.0 and 5.0 mbar) in the stratosphere.  \added{Assuming geostrophic balance (a valid assumption at polar latitudes with Rossby numbers less than 0.2\cite{14sanchez}),} meridional temperature gradients ($\partial{T}/\partial{y}$, where $y$ is the north-south distance) are in thermal \added{wind} balance with the vertical shear on the zonal winds\cite{04holton} ($\partial{u}/\partial{z}$, where $u$ is the zonal wind and $z$ is altitude).  This allows us to \added{estimate} the windshear $\partial{u}/\partial{z} =-(g/fT)\partial{T}/\partial{y}$, where $g$ is the gravitational acceleration for each latitude and altitude, and $f$ is the Coriolis parameter.  This provides us with a means of expressing the changing meridional gradients as a function of time, where the windshear can be interpreted as enhancing eastward winds (positive shear) or westward winds (negative shear) if a level-of-zero-motion exists somewhere in the lower stratosphere. Tropospheric temperatures and windshear are shown in the Supplementary Figure 1.

Fig. \ref{stratos_temp} allows us to compare the disappearance of the warm SPSV to the onset of the NPSV.  Warm stratospheric polar vortices become evident when \added{strong temperature gradients develop over time, strengthening negative windshear to promote westward winds.}. The zonal mean temperatures reveal latitudinal gradients that are strongest at 5 mbar, reminiscent of the banded structure that dominates in the troposphere.  These contrasts are \added{not as apparent} at 0.5-1.0 mbar, where temperatures vary more smoothly with latitude and are dominated by negative shear poleward of the prograde zonal jets at $78^\circ$N and $74^\circ$S.  At the start of the Cassini time series, the SPSV had two primary structures: the negative $dT/dy$ near $88^\circ$S that defines the edge of the compact polar cyclone\cite{05orton}, and the negative $dT/dy$ near $74^\circ$S that represents the edge of the SPSV\cite{08fletcher_poles} (Fig. \ref{stratos_temp}j-l).  In 2006 ($L_s=320^\circ$) temperatures increased towards the South Pole at all altitudes, but in 2007-2008 a cool band began to form at 5 mbar centred on $84^\circ$S, signified by the onset of the positive windshear in Fig. \ref{stratos_temp}g-j.  \added{By 2013-2014 ($L_s\approx40^\circ$), the negative shear at the SPSV boundary ($74^\circ$S) had weakened considerably and a cool band was beginning to form near $80-85^\circ$S at 1 mbar, signifying the demise of the SPSV.  The cooling and dissipation of the SPSV started pre-equinox ($L_s\approx340^\circ$), as measured from the timing of the most negative windshear at $74^\circ$S.}

\added{The latitudinal structure of the vertical} windshear in the northern hemisphere (Fig. \ref{stratos_temp}d-f) is remarkably similar to that in the south, with higher pressures exhibiting banded temperature structures and lower pressures exhibiting smooth latitudinal gradients.  Negative windshear dominates poleward of $78^\circ$N and $p<1$ mbar throughout the time series, because temperatures have always increased towards $90^\circ$N despite the sequence starting in polar winter, counter to the expectations of radiative climate models\cite{14guerlet, 15fletcher_poles, 16hue}.  A 5-mbar region of positive windshear (Fig. \ref{stratos_temp}c,f, i.e., a cool band, mirroring the one that formed in the SPSV) was centred on $83^\circ$N until 2017, when it had weakened and was replaced by warmer temperatures (i.e., 5-mbar temperatures increased everywhere poleward of $\sim75^\circ$N \added{by the end of the mission}).  Similar positive windshear bands, centred over the broad westward tropospheric jets, existed at $73\pm1^\circ$N and $63\pm1^\circ$N, \added{but these all weakened as Saturn approached solstice, indicating a positive $dT/dy$ gradient throughout the north polar region}.  \added{In particular,} the temperatures near the $78^\circ$N eastward jet in Fig. \ref{stratos_temp}a \added{exhibited} a strengthening $\partial{T}/\partial{y}$ gradient with time, signifying the development of the NPSV boundary. The windshear on this jet transitioned from weakly positive to negative at 5 mbar in 2012-13 ($L_s\approx40^\circ$), \added{which would serve to promote westward winds surrounding the summertime vortex.}  

However, by the end of the mission near northern summer solstice ($L_s=93^\circ$), the NPSV boundary windshear was less than half of that at the SPSV boundary in 2006-2008, and the 5-mbar temperature maximum ($\sim140$ K at $90^\circ$N in Fig. \ref{stratos_temp}c) was still cooler than that the maximum temperature at the south pole ($\sim150$ K at $90^\circ$S in Fig. \ref{stratos_temp}l).  So although the NPSV had been developing a strong thermal gradient since the end of the last CIRS study in northern spring\cite{15fletcher_poles}, and positive windshear zones had disappeared by northern summer, it still lacked the contrasts exhibited by the SPSV \added{during the later parts of} southern summer.  \added{As the thermal maximum lags behind the seasonal insolation maximum\cite{15fletcher_poles}, we expect that the NPSV continued to warm beyond northern summer solstice and the end of the Cassini observations.}

\subsection{Composition of the NPSV and SPSV.}

\begin{figure}
\centering
\makebox[\textwidth][c]{\includegraphics[width=1.3\textwidth]{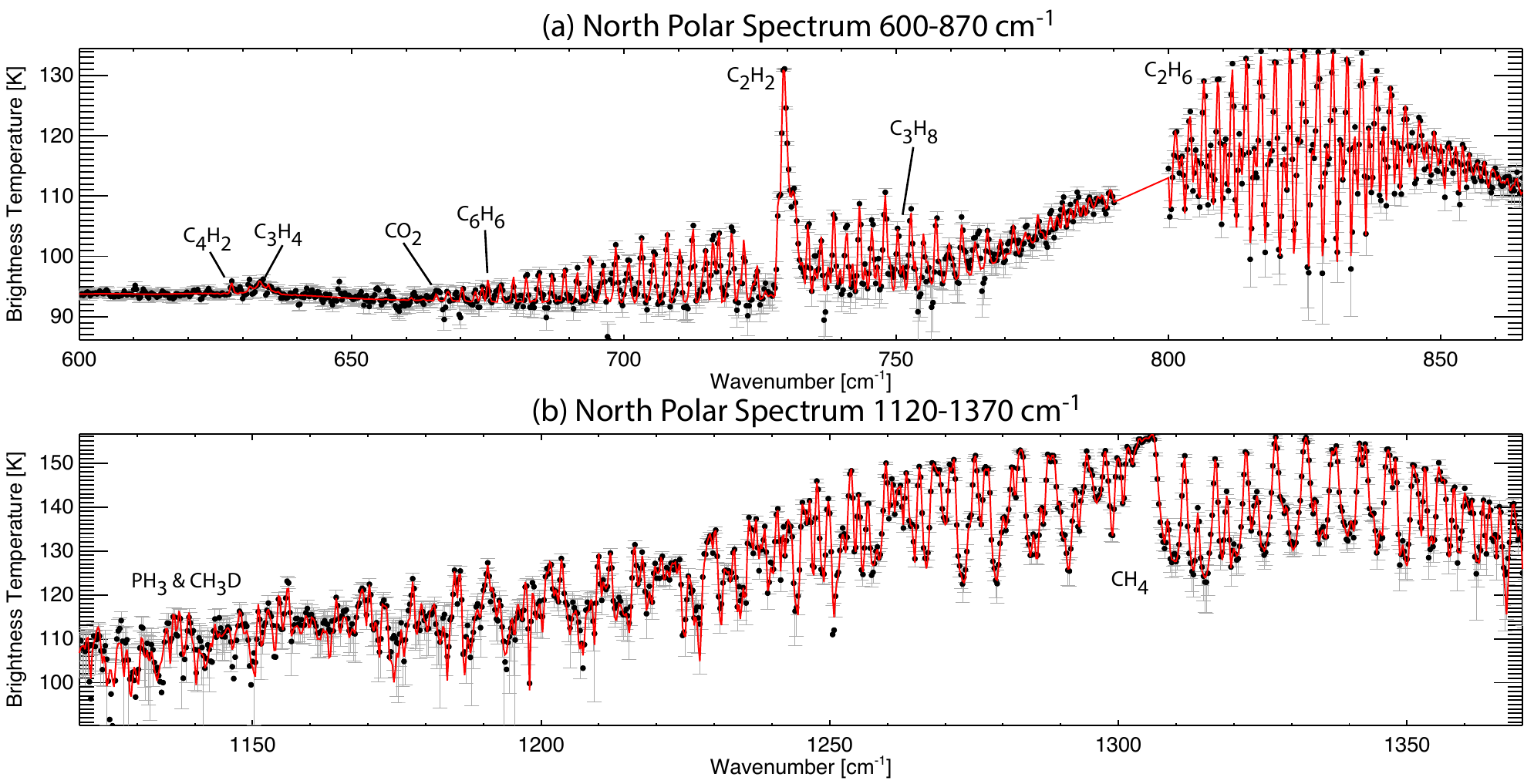}}
\caption{\textbf{Best-fitting brightness temperature spectrum for the 0.5-cm$^{-1}$ resolution observation of the NPSV on March 27th 2017 (spanning $84-87^\circ$N)}.  The figure shows the model fit (red) to the data (black points with grey uncertainties).  The figure is split into the two spectral ranges used in the inversions: (a) 600-870 cm$^{-1}$ and (b) 1120-1370 cm$^{-1}$.  Key gaseous features are labelled.  Data between 870 and 1120 cm$^{-1}$ were omitted from the inversion.  The measurement uncertainties (grey bars) depend on the number of coadded CIRS spectra and the number of deep space reference spectra used in the calibration.}
\label{compsit_spx}
\end{figure}

Although the NPSV was forming in the final years of Cassini's mission, the thermal gradients (and negative windshear) have not yet reached the extent of those observed in the southern hemisphere surrounding the SPSV.  This may not be unexpected - northern summer occurred near aphelion (April 17, 2018) whereas southern summer was closer to perihelion (July 26, 2003), implying a difference in the rate of energy deposition in the upper atmosphere.  However, CIRS offers another technique for assessing the similarities and differences between the NPSV and SPSV by studying the relative enhancements of different chemical species within each vortex.  This requires fitting high-spectral resolution CIRS observations that stared directly at each vortex (see Methods), as shown for the NPSV in Fig. \ref{compsit_spx}.

\begin{figure}
\centering
\makebox[\textwidth][c]{\includegraphics[width=0.9\textwidth]{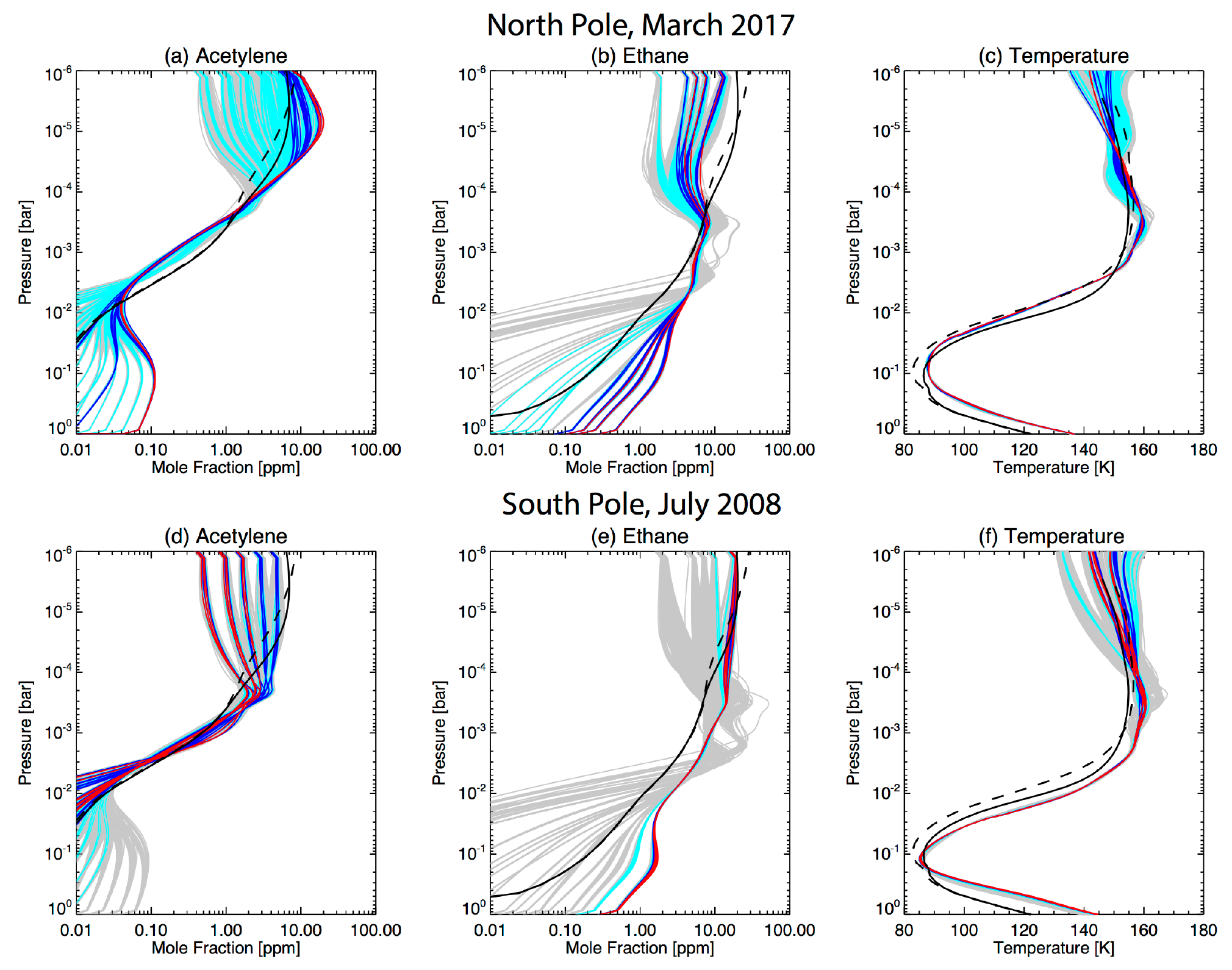}}
\caption{\textbf{Vertical distributions of temperature, acetylene and ethane derived from 0.5-cm$^{-1}$ spectral resolution data of both poles.} The panels show COMPSITs targeting the north pole in March 2017 (panels a-c, $84-87^\circ$N) and the south pole in July 2008 (panels d-f, $83-88^\circ$S).  A grid of 2000 \textit{a priori} atmospheres (varying the temperature and hydrocarbon distributions) was employed as starting point.  Red lines indicate profiles fitting the spectrum within $1\sigma$ of the best fit, blue lines are for $2\sigma$, turquoise lines are for $3\sigma$ and grey lines are for optimal estimates that do not fit within the $3\sigma$ envelope. We have confidence in retrieved profiles in altitude ranges where the profiles converge, irrespective of the \textit{a priori} starting points.  The hydrocarbon profiles are compared to the output of a diffusive photochemistry model\cite{07moses} at $\pm72^\circ$ latitude for $L_s=88^\circ$ (dashed black line) and $L_s=346^\circ$ (solid black line).  The temperature profiles are compared to the predictions or a radiative climate model\cite{14guerlet} in the absence of stratospheric aerosols for $\pm85^\circ$ latitude and for $L_s=88^\circ$ (dashed black line) and $L_s=346^\circ$ (solid black line). }
\label{priortest}
\end{figure}

Vertical profiles of temperature, ethane, and acetylene were derived starting from a broad ensemble of priors (see Methods), and are displayed in Fig. \ref{priortest} for both the NPSV ($L_s=88^\circ$) and the SPSV ($L_s=346^\circ$).  The temperature profiles are compared with the predictions of a radiative climate model\cite{14guerlet} for the same season, \added{which assumed a time-invariant low-latitude mean of the stratospheric coolants ethane and acetylene derived from CIRS limb observations between 2005 and 2008\cite{09guerlet}}.  Whilst the northern $T(p)$ is consistent with these predictions for 2-100 mbar, the southern $T(p)$ is always warmer, even though the maximum stratospheric temperatures ($T\sim160$ K at 0.3-0.5 mbar) are similar at both poles.  This hints at a missing source of heat \added{in the radiative model for} the lower stratosphere during southern summer that was not required in the north at $L_s=88^\circ$, maybe due to stratospheric aerosols (radiative) or stronger polar subsidence (adiabatic heating)\cite{14guerlet, 15fletcher_poles, 15guerlet}, \added{the latter being supported by the excess C$_2$H$_2$ observed in the SPSV compared to the NPSV in Table \ref{tab:compsit_abundance}, discussed below.}  At lower pressures, both the NPSV and SPSV inversions suggest a weak stratopause, with temperatures declining from 160 K at 0.3 mbar to $\sim150$ K at 0.01 mbar, again consistent with the radiative model with and without the presence of polar aerosols\cite{15guerlet}.  Although this is dependent on the prior and hampered by the nadir viewing geometry, it is worth noting that \added{this decrease in temperature occurs} even with an isothermal 170-K prior.  Intriguingly, the presence of a mesosphere overlying a warmer stratosphere was also inferred from ground-based IRTF observations of Saturn's southern hemisphere\cite{05greathouse_dps} at $L_s=268^\circ$, before Cassini's arrival.  

The vertical hydrocarbon profiles in Fig. \ref{priortest} are less well constrained than the temperatures, but they show that CIRS 0.5 cm$^{-1}$ spectral resolution data can constrain the acetylene gradient between 0.05 and 5 mbar and the ethane gradient between 0.5 and 5 mbar.  Outside of these ranges, the profiles diverge and tend back to their respective priors.  For this reason, abundances at low pressures ($p<0.1$ mbar) should be treated with caution, but there is a trend for the south polar data to exhibit well-mixed hydrocarbons with altitude (i.e., no deviation from the uniform prior at low pressures), whereas north polar ethane shows a slight decline with altitude, and north polar acetylene shows a high-altitude peak near 0.01 mbar.  Ethane is enhanced in the 0.5-5 mbar range compared to the expectations of photochemical models\cite{07moses, 15hue, 16hue} for both poles, and exhibits a somewhat shallower vertical gradient than predicted.  This is qualitatively consistent with the idea of a long-lived photochemical product being advected downwards over the pole.  Conversely, the acetylene gradient is steeper than expected from the photochemical model, suggesting photochemical depletion at a rate that is faster than the downward advection.  The 1-mbar acetylene abundance is well-matched at the south pole, but somewhat depleted at the north pole compared to predictions.  

\subsection{Minor hydrocarbons.}

\begin{figure}
\centering
\makebox[\textwidth][c]{\includegraphics[width=1.3\textwidth]{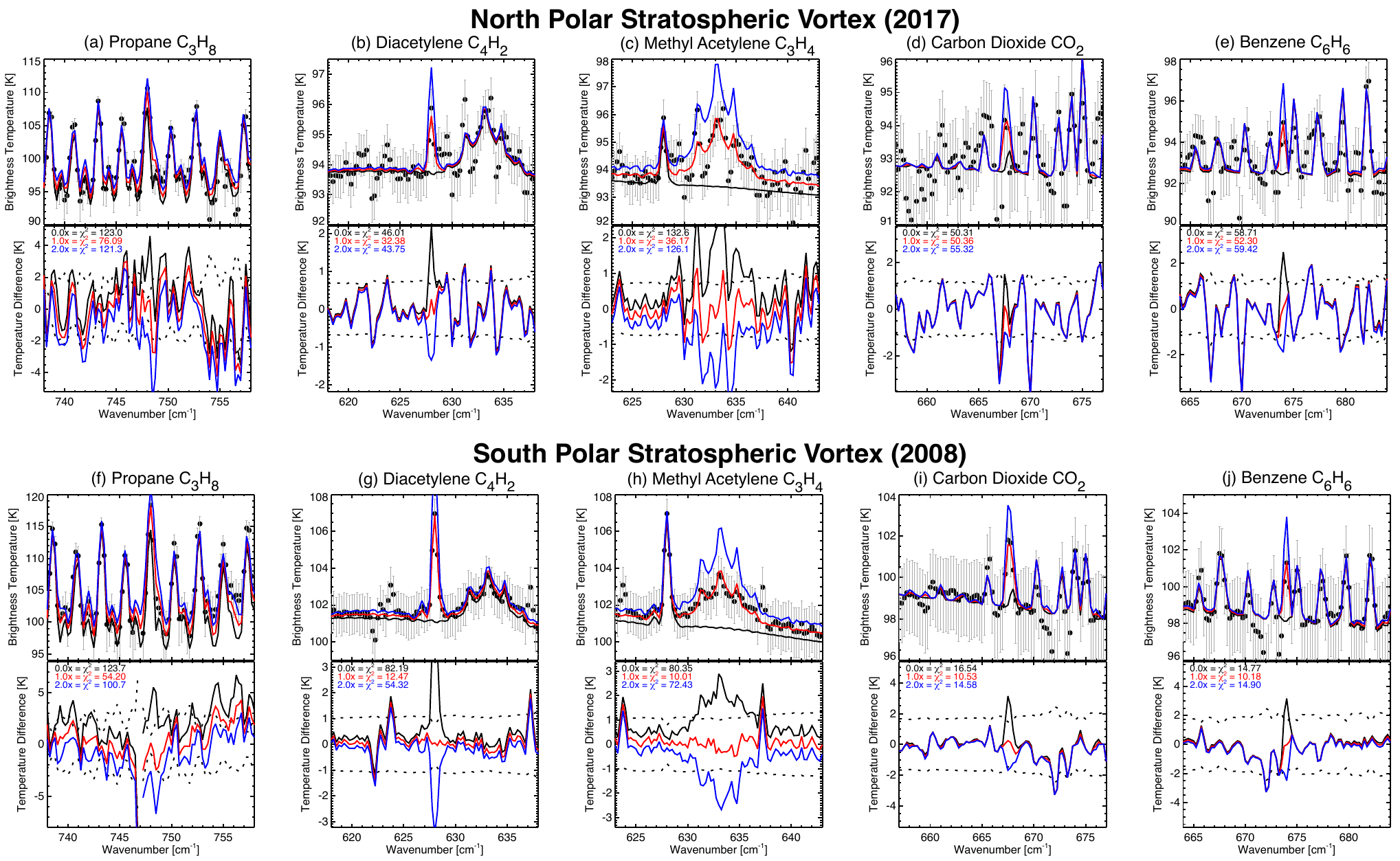}}
\caption{\textbf{Detection of minor species within Saturn's northern and southern polar vortices.}  The NPSV was observed in March 2017 (panels a-e), the SPSV was observed in July 2008 (panels f-j), using the same data as in Fig. \ref{priortest}. Five minor species are shown:  propane (panels a and f), diacetylene (panels b and g), methyl acetylene (panels c and h), carbon dioxide (panels d and i), and benzene (panels e and j).  Once the best-fitting atmospheric profiles (temperatures and hydrocarbons) had been established, we scaled the abundance of five minor species by 0.0 (black), 1.0 (red) and 2.0 (blue) times to show their detectability in the spectrum.  For each panel and gaseous species, we show the brightness temperature spectrum and its uncertainty, as well as the difference spectrum between the data and the model.  The goodness-of-fit is reported for each model, with a $\Delta \chi^2>1$ signifying a $1\sigma$ detection. }
\label{cxhy_minor}
\end{figure}

Adopting the best-fitting \added{(i.e., minimum $\chi^2$)} temperature, ethane and acetylene profiles from Fig. \ref{priortest}, the 0.5-cm$^{-1}$ spectra were used to search for evidence of minor species in both the NPSV and SPSV \added{by scaling the model profiles\cite{07moses} of each species during the spectral inversion}.  Fig. \ref{cxhy_minor} shows the effect on the spectra of removing a gas species or doubling its best-fitting abundance, compared to the uncertainty on the coadded CIRS spectrum.  The change in $\chi^2$, evaluated over the spectral range shown in each figure, is given beneath each brightness temperature spectrum.  Methyl-acetylene and diacetylene are both clearly detected at both poles, whereas detections of propane, CO$_2$ and benzene are more marginal, particularly in the NPSV. 

The retrieved abundances are recorded in Table \ref{tab:compsit_abundance}, and show that the ratio of the NPSV and SPSV abundances are always smaller than one, contrary to the expectations of a photochemical model\cite{07moses} for the same season.  This provides another indication that the NPSV was not as mature as the SPSV when viewed at northern summer solstice.  Only ethane and benzene have abundances that are consistent between both poles \added{(although with the largest uncertainties)}, whereas all other species are less abundant by 20-60\% in the north compared to the south.  The uncertainties on the mole fractions in Table \ref{tab:compsit_abundance} are the uncertainties on the scale factor for the prior abundance - propane and diacetylene exceed the expectations of the photochemical model\cite{07moses}, methyl-acetylene is close to expectations, and CO$_2$ and benzene are smaller than expectations.  That benzene is $0.6-0.8\times$ less than 1-mbar photochemical predictions at the poles is consistent with previous findings using limb-sounding CIRS spectra\cite{15guerlet}.  Nevertheless, the measured ratio deviates from the model ratio\cite{07moses} substantially for all but ethane and benzene.  

There may be a weak trend with the longevity of the photochemical species\cite{05moses_sat}, with longer-lived ethane and propane showing the most similarity between the poles (ratios of 0.7-1.0), and the shorter-lived acetylene, diacetylene and methyl-acetylene showing greater contrasts (0.2-0.6).  This makes sense if the shorter-lived species are more sensitive to the dynamical and chemical changes associated with the polar vortex formation, although this qualitative view ignores any potential feedbacks between dynamics, chemistry and radiative cooling in the vortices.  The challenge to future seasonal photochemistry models is to reproduce the observed gradients by \added{better constraining} the vertical and horizontal mixing rates.  

\begin{table}
\caption{Abundances of gases in NPSV and SPSV at 1 mbar [ppbv] compared to the expected ratio from a photochemical model\cite{07moses} for the same season.  With the exception of ethane and acetylene, these values were derived by scaling the \textit{a priori} profiles from Moses et al.\cite{07moses}.}
\centering
\begin{tabular}{l c c c c}
\hline
Gas  & NPSV & SPSV & Ratio (N/S) & Model Ratio \\
\hline
Acetylene C$_2$H$_2$ & $225\pm65$ & $423\pm132$ & $0.53\pm0.23$ & $1.04$\\
Ethane C$_2$H$_6$ & $8000\pm1841$ & $8214\pm4585$ & $0.97\pm0.58$ & $0.98$ \\
Methyl-Acetylene C$_3$H$_4$ & $2.0\pm0.2$ & $3.3\pm0.5$ & $0.62\pm0.11$ & $0.91$ \\
Propane C$_3$H$_8$ & $100\pm16$ & $147\pm20$ & $0.68\pm0.14$ & $0.96$ \\
Diacetylene C$_4$H$_2$ & $0.11\pm0.03$ & $0.53\pm0.07$ & $0.20\pm0.06$ & $1.10$ \\
Benzene C$_6$H$_6$ & $0.26\pm0.10$ & $0.33\pm0.14$ & $0.80\pm0.45$ & $0.95$ \\
Carbon Dioxide CO$_2$     & $0.2\pm0.1$ & $0.5\pm0.2$ & $0.37\pm0.26$ & $1.02$ \\
\hline
\label{tab:compsit_abundance}
\end{tabular}
\end{table}

\subsection{Chemical distributions within the NPSV.}

\begin{figure}
\centering
\makebox[\textwidth][c]{\includegraphics[width=1.3\textwidth]{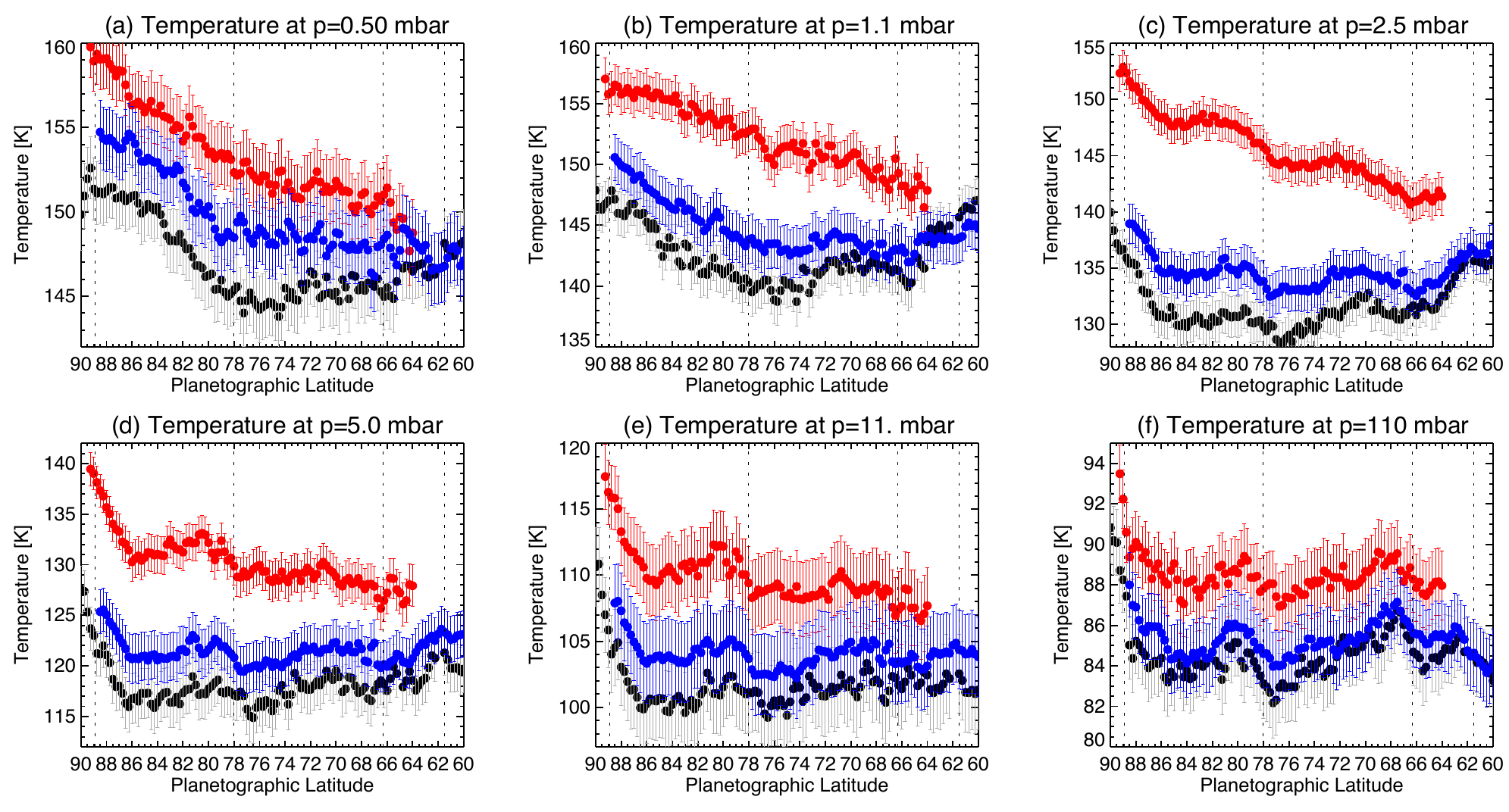}}
\caption{\textbf{Zonal-mean north polar temperatures for three REGMAP datasets showing the warming of the polar stratosphere.}  Temperatures are displayed at six pressure levels:  (a) 0.5 mbar, (b) 1.1 mbar, (c) 2.5 mbar, (d) 5.0 mbar, (e) 11 mbar, and (f) 110 mbar. Three dates are included in each panel:  August 17th 2013 (black, REGMAP $196\_001$), June 25th 2014 (blue, REGMAP $205\_001$) and February 12th 2017 (red, REGMAP $261\_003$).  Vertical dotted lines show the locations of prograde jet peaks\cite{15antunano} for comparison with the meridional temperature gradients.  Error bars represent the formal retrieval uncertainty.  Note that the June 2014 dataset utilised 15-cm$^{-1}$ resolution spectra, whereas the 2013 and 2017 datasets are better constrained by 2.5-cm$^{-1}$ resolution spectra. }
\label{comp_T}
\end{figure}

Despite the strong constraints on abundances provided by the 0.5-cm$^{-1}$ CIRS spectra, they lack the spatial coverage to track the latitudinal variation of these species.  A compromise is to consider the variations of only temperature, ethane and acetylene using the 2.5-cm$^{-1}$ resolution REGMAP observations.  The north polar priors (Fig. \ref{priortest}a-c) were used as the starting point for latitudinally-resolved retrievals between 2013 and 2017 shown in Fig. \ref{comp_T}. These figures reinforce the warming trends described above, and show the difference between $p<1$ mbar, where the $\partial{T}/\partial{y}$ gradients are relatively smooth with latitude, and $p>2$ mbar, where the banded structure that typifies the troposphere can be seen.  The north polar cyclone is easily observed on all dates as the sharp temperature rise within $3^\circ$ of the pole for $p>1$ mbar, but is not evident for $p<1$ mbar.  The 2.5-, 5- and 10-mbar temperature gradients are particularly strong across the latitude of the hexagonal jet at $78^\circ$N, which is why the stratospheric hexagon appears readily in Fig. \ref{hexmaps}.  

\begin{figure}
\centering
\makebox[\textwidth][c]{\includegraphics[width=1.3\textwidth]{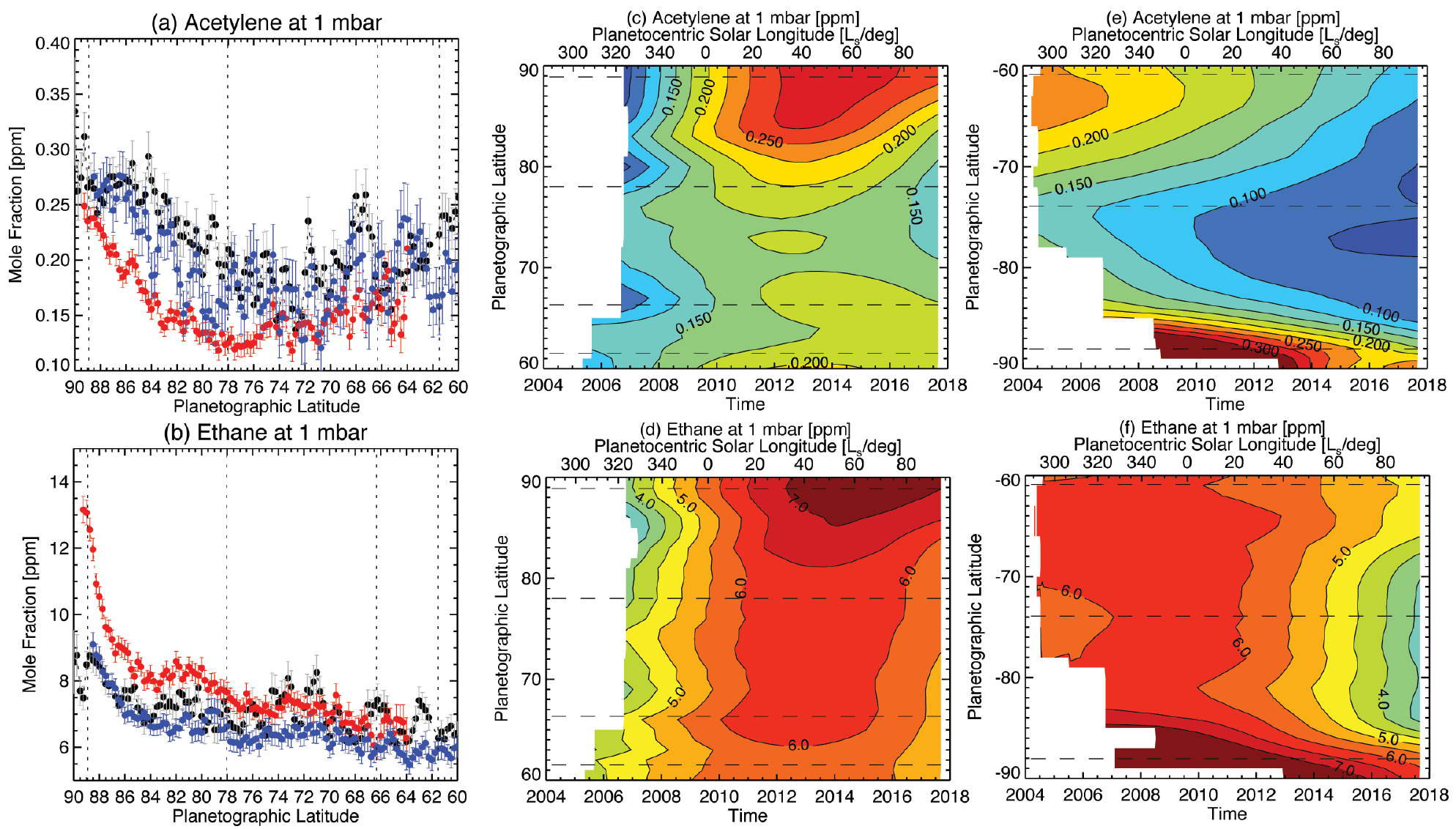}}
\caption{\textbf{Temporal evolution of polar ethane and acetylene.} Latitudinal distributions of (a) acetylene and (b) ethane at 1 mbar as derived from spectra on August 17th 2013 (black, REGMAP $196\_001$), June 25th 2014 (blue, REGMAP $205\_001$) and February 12th 2017 (red, REGMAP $261\_003$). These results come from scaling of the hydrocarbon profiles derived for the north pole from the 0.5-cm$^{-1}$ resolution observations, at the same time as retrieving the temperature profiles in Fig. \ref{comp_T}.  Uncertainties on these retrieved scale factors are shown for each data point.  The four contour plots show the spline-interpolated results for (c) north polar aceylene, (d) north polar ethane, (e) south polar acetylene, and (f) south polar ethane from the low-resolution time series in Fig. \ref{stratos_temp}.  Dotted/dashed lines show the locations of prograde jet peaks\cite{15antunano} in all six panels.}
\label{comp_gas}
\end{figure}

The 1-mbar distributions of acetylene and ethane at the North Pole for the same three dates are shown in Fig. \ref{comp_gas}a-b, and were generated by scaling the best-fitting vertical profiles in Fig. \ref{priortest}.  We caution the reader that these simple scalings of the \textit{a priori} gas abundances would omit any changes in the vertical hydrocarbon gradients, which are assumed to be invariant in the present study.  These are compared to the 1-mbar abundances at both poles derived from the low-resolution inversions, which have been smoothed and interpolated via a tensioned spline\cite{07teanby_splines,15fletcher_poles}.  C$_2$H$_2$ shows a minimum near $75^\circ$N and increases towards the poles by a factor of $\sim2$ (Fig. \ref{comp_gas}c), similar to that seen poleward of $75^\circ$S when the SPSV was present (Fig. \ref{comp_gas}e).  However, whilst there was little change between 2013-2014, there has been a surprising drop in C$_2$H$_2$ by 2017 which is counter to the pre-2014 trend\cite{15fletcher_poles}.  This reduction is seen in both the 2.5-cm$^{-1}$ and 15-cm$^{-1}$ inversions, suggesting that the C$_2$H$_2$ abundance peaked before summer solstice.  However, the comparison with the SPSV suggests that the NPSV has not reached the same elevated abundance during this time series, as suggested in Table \ref{tab:compsit_abundance}.

The distribution of C$_2$H$_6$ is different, showing an overall increase in the northern hemisphere and a potential spike in the abundance poleward of $86^\circ$N (Fig. \ref{comp_gas}b), although this is not seen in the low-resolution time series (Fig. \ref{comp_gas}d).  A similar high-latitude peak in ethane may have been present in the centre of the SPSV early in the Cassini mission (Fig. \ref{comp_gas}f).  In summary, both C$_2$H$_2$ and C$_2$H$_6$ increase poleward of the NPSV boundary at $78^\circ$N, similar to chemical enhancements observed in the SPSV at the start of the Cassini mission.  Such an increase is not seen in photochemical models that lack meridional motions\cite{07moses, 15hue}, and is further evidence that subsidence is occurring throughout the NPSV \added{to generate the warming and elevated chemical abundances\cite{16sayanagi}}.  Based on the record of the SPSV, we would expect these polar enhancements to start to decline after the summer solstice ($L_s=90^\circ$).  However, we caution that ion chemistry in the auroral regions, which could also be modulating the hydrocarbon abundances, remains to be explored in detail.

\section*{Discussion}

Our results indicate that Saturn's long-awaited NPSV started to develop during late northern spring ($L_s>40^\circ$), but that temperature gradients only became positive poleward of $78^\circ$N at 5 mbar in 2017, around northern summer solstice ($L_s=90^\circ$).  Even so, neither the temperature gradients, nor the peak temperatures, nor the stratospheric molecular abundances, were able to match those observed in the SPSV in late southern summer.  Either the NPSV will never reach the same contrasts as those of the SPSV, or the NPSV will continue to develop through early northern summer.  

Most surprisingly, the boundary of the NPSV \added{exhibited} a hexagonal shape mirroring that observed in the clouds $\sim300$ km below.  Saturn's polar hexagon was discovered by Voyager\cite{88godfrey}, re-observed during Saturn's last northern summer in the early 1990s by Hubble\cite{93caldwell} and ground-based facilities\cite{93sanchez}, and re-detected by Cassini in the thermal-infrared\cite{08fletcher_poles, 15fletcher_poles}, near-infrared\cite{09baines_pole} and visible\cite{14sanchez, 15antunano, 17sayanagi, 18antunano, 18requena}.  The meandering of the jetstream that forms the hexagon is believed to be a Rossby wave\cite{90allison_hex} resulting from an instability of the eastward zonal jet near $78^\circ$N\cite{09read_epv, 10barbosa, 14sanchez, 15morales} and trapped within a waveguide formed by the zonal jets.  

\begin{figure}
\centering
\makebox[\textwidth][c]{\includegraphics[width=1.3\textwidth]{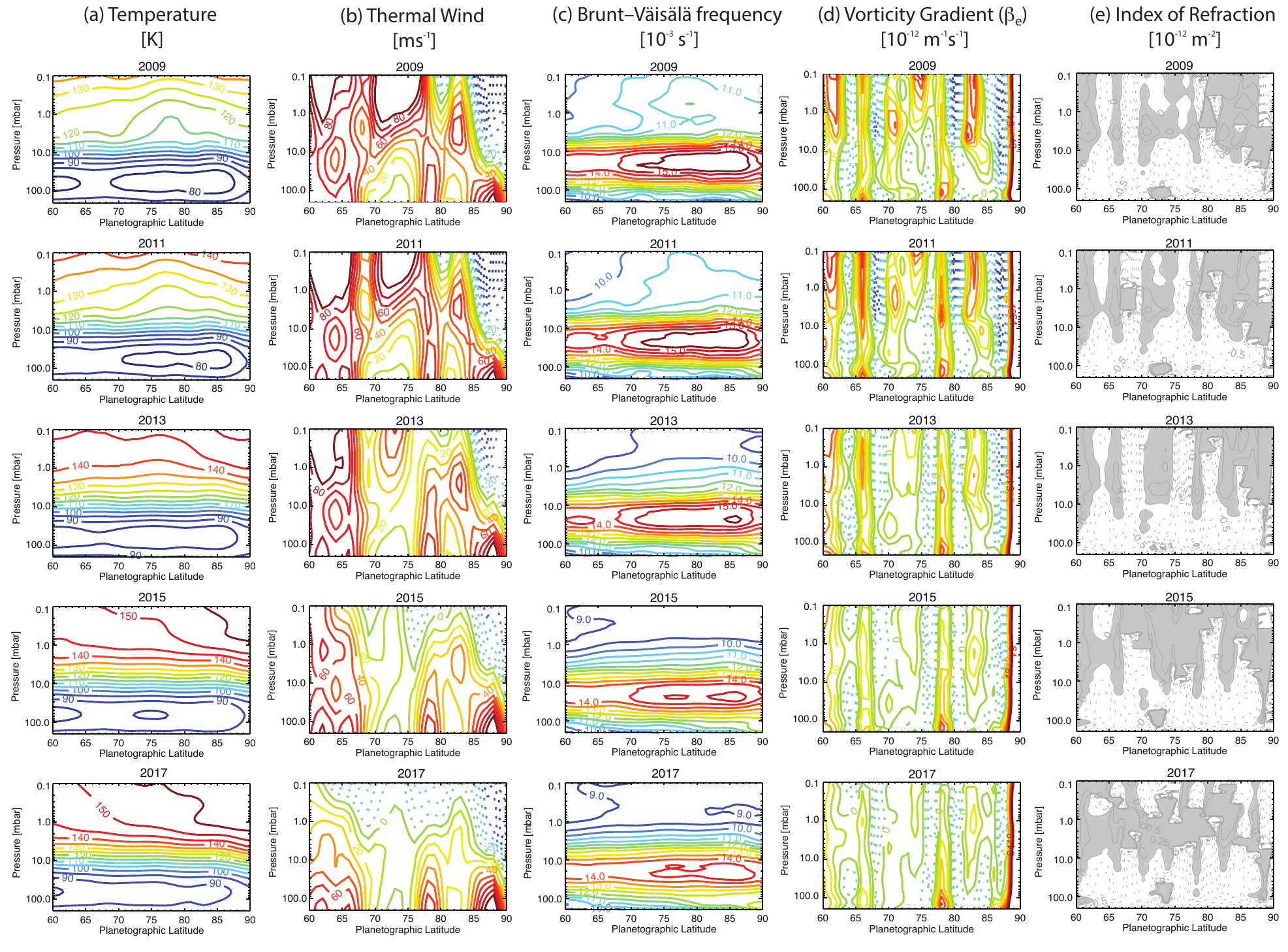}}
\caption{\textbf{Derived dynamical products based on the reconstructed north polar time series.}  Annual averages for 2009, 2011, 2013, 2015 and 2017 are shown in the five rows.  We show (a) the temperatures $T$, (b) zonal winds $u$, (c) Brunt V\"{a}is\"{a}l\"{a} buoyancy frequency $N$, (d) effective vorticity gradient $\beta_e$, and (e) the index of refraction $\nu^2$.  Negative values are shown as dotted contours, positive values as solid contours.  In panel (e), regions of real refractive index are shown as grey.  In panel (d), the scale for $\beta_e$ ranges from $\pm80\times10^{-12}$ m$^{-1}$s$^{-1}$ in steps of $8\times10^{-12}$ m$^{-1}$s$^{-1}$.  Contour lines in panel (e) for $\nu^2$ are drawn for $\pm0.5\times10^{-12}$ m$^{-2}$,  $\pm0.25\times10^{-12}$ m$^{-2}$ and 0 m$^{-2}$.  Given the uncertainties in the thermal winds (see Methods), these panels should be considered as qualitative trends rather than quantitative values.}
\label{Npole_dynamics}
\end{figure}

Fig. \ref{Npole_dynamics} shows how the 2D temperatures and winds derived from the CIRS time series can be used to explore the nature of the wave and its vertical propagation.  \added{Mean} zonal winds $u$ are calculated via the thermal wind equation\cite{04holton}, using the latitudinal temperature gradients of Fig. \ref{stratos_temp} to integrate the cloud-top winds\cite{15antunano} (assumed to reside at 500 mbar).  Gradients in \added{temperatures, winds, and vorticity} provide a valuable diagnostic tool for understanding wave propagation, and we follow previous work\cite{16fletcher} in estimating the latitudinal gradient of quasi-geostrophic potential vorticity ($\beta_e=\partial{q_G}\partial{y}$), where $y$ is the north-south distance and $q_G$ is the potential vorticity\cite{87andrews}. $\beta_e$ is the sum of three terms:  (1) $\beta=\partial{f}/\partial{y}$, the northward gradient of planetary vorticity (i.e., the Coriolis parameter $f$); (2) $\beta_y=-\partial^2u/\partial{y^2}$, the meridional curvature of the zonal wind field; and (3) $\beta_z$, related to the vertical curvature of the wind field and the static stability.  This reduces to\cite{16fletcher}:
\begin{eqnarray}
\beta_e=\beta+\beta_y+\beta_z \\
\beta_e=\beta-\frac{\partial^2u}{\partial y^2}-\frac{1}{\rho}\pderiv{}{z}\left(\rho \frac{f^2}{N^2}\pderiv{u}{z}\right)
\end{eqnarray}
Here, \added{$\rho$ is the density} and $N$ is the Brunt V\"{a}is\"{a}l\"{a} frequency:
\begin{equation}
N^2=\frac{g}{T}\left(\pderiv{T}{z} + \frac{g}{c_p}\right)
\end{equation}
The latitudinal and vertical distribution of $N$ is shown in Fig. \ref{Npole_dynamics}c, calculated using the spatial distribution of gravitational acceleration $g$ assuming an oblate spheroid; and the specific heat capacity $c_p$ based on an equilibrium ratio of ortho-to-para-hydrogen \added{for all dates in the time series, and including contributions from helium and methane.}   

The resulting $q_G$ gradient is shown in Fig. \ref{Npole_dynamics}d, revealing that the hexagon is co-located with a narrow lane of positive $\beta_e$ in both the troposphere and stratosphere, and resides between two latitudes where $\beta_e$ changes sign.  This reinforces the idea that the vorticity gradient serves as a waveguide for the propagating Rossby wave\cite{14sanchez}, suggestive of an atmospheric `duct' where the hexagon is observed.  The $\beta_e$ sign reversal suggests that the atmosphere violates both the neutral-stability criterion relative to Arnol'd's second theorem\cite{09read_epv}, and the Charney-Stern\cite{62charney} and Rayleigh-Kuo necessary conditions for baroclinic and barotropic stability, respectively. Although the possibility of instability is not sufficient to explain the hexagon's origins, waves and large-scale eddies are often found in such locations\cite{09read_epv}.  

The ability of a Rossby wave to propagate \added{is related to} the index of refraction ($\nu^2$) of the background medium\cite{61charney}, which is the sum of the squares of the vertical $m$, meridional  $l$, and zonal $k$ wavenumbers\cite{96achterberg, 16fletcher}:
\begin{equation}
\nu^2 \equiv k^2+l^2+\left(\frac{f^2}{N^2}\right)m^2
\end{equation}
This can be related to the dispersion relationship for a 3D Rossby wave in a baroclinic atmosphere\cite{11sanchez}:
\begin{equation}
\nu^2 = \frac{\beta_e}{u-c_x} - \frac{f^2n^2}{N^2}
\end{equation}
where $n^2$ includes the higher-order vertical derivatives of the buoyancy frequency\cite{96achterberg, 16fletcher}.  This index depends only on the phase speed of the wave ($c_x$) and the properties of the background medium ($u$, $N^2$, scale height $H$, $f$ and $\beta_e$).  \added{With known values of $k$ and $l$, this can be rearranged to estimate the vertical wavenumber $m$: $m^2>0$ implies that Rossby waves are permitted to propagate in the vertical, whereas regions of $m^2<0$ imply evanescent waves that decay exponentially with altitude.}  

\added{To investigate whether changing atmospheric conditions have permitted the tropospheric hexagon to propagate into the stratosphere,} Fig. \ref{Npole_dynamics}e shows the subtle changes in $\nu^2$ since spring equinox, calculated assuming the \added{measured phase speed of the tropospheric hexagon} $c_x=-0.036$ ms$^{-1}$\cite{14sanchez}, and confirms that regions of positive $\nu^2$ are confined to low pressures by the static stability (the $\beta_z$ term, which changed with time), and confined latitudinally by the curvature of the windfield (the $\beta_y$ term, which was relatively constant over time).  The hexagon latitude is characterised by a vertically-extended region of positive $\nu^2$ in the stratosphere coincident with a region of positive vorticity gradient $\beta_e$ between $77-80^\circ$N, which is suggestive of a wave duct that can be seen to extend downwards as northern solstice approached (i.e., the static stability changed with season), but never quite reached the tropopause. These distributions are subject to large uncertainties\cite{16fletcher}, most notably because the temperatures and winds in the 5-80 mbar range are poorly constrained by the CIRS inversions (see Methods), \added{such that quantitative values should be treated with caution.}

\added{However, taking values of $k=2\pi/L_x$ (where $L_x=$14,500 km is the length of the hexagon's side\cite{14sanchez}) and $l=2\pi/L_y$ (where $L_y=3150$ km, estimated as the distance between the two $\beta_e$ sign changes at $77-80^\circ$N in Fig. \ref{Npole_dynamics}d, which could demarcate the latitudinal boundaries of the wave), we searched for regions where vertical propagation is possible for a hexagon-like Rossby wave (i.e., where $m^2>0$).  Although a region of positive $m^2$ does exist for the hexagon at $p>600$ mbar (i.e., near the cloud tops), we found $m^2<0$ throughout the upper troposphere and stratosphere, supporting the `trapped-wave' concept.  This poses a problem for the hexagon observed in both the upper troposphere ($p\sim100$ mbar) and stratosphere ($p\sim1$ mbar).}

At least two possibilities should therefore be considered.  Firstly, if the magnitude of the imaginary $m$ were sufficiently small, any vertically-propagating wave could travel by evanescence (tunnelling) over a limited altitude range from the troposphere into the stratosphere.  Although the coupling of these two altitude ranges may be weak, the amplitude of a Rossby wave will still tend to increase with height because of the decreasing density of the air.  The fact that we do not see any zonal phase shift of the hexagon vertices between the upper troposphere and stratosphere (Fig. \ref{vertices}) is consistent with the idea that the wave is not formally propagating.  \added{Secondly, the $\beta_e$ sign reversal in Fig. \ref{Npole_dynamics}d shows that the atmosphere exhibits the necessary conditions for instability in the stratosphere, irrespective of what is happening in the troposphere.  However, if such a pattern were excited independently, one would need to explain why it retains the same wavenumber, meridional amplitude, and phase speed as its tropospheric counterpart.  Such a stable phase alignment would require some form of dynamical coupling between these two altitude ranges.}  Distinguishing between these cases will require investigation via more detailed middle atmosphere modelling.  Finally, we note that we found no evidence of stratospheric polygonal waves surrounding the SPSV during southern summer, consistent with their absence in the troposphere, and confirming the differences in dynamical stability of Saturn's northernmost ($78^\circ$N) and southernmost ($74^\circ$S) eastward jets.




Finally, we discuss potential sources of the heating responsible for the formation of the NPSV.  Supplementary Figure 2 compares the CIRS temperatures at 1 mbar to the results of radiative models from several different studies.  Although they differ in the details, each model qualitatively reproduces the amplitude of the temperature changes and the timing of the temperature minima/maxima.  However, none of the models successfully reproduces the formation of $\partial{T}/\partial{y}$ gradients associated with the edges of the compact polar cyclones or the broad stratospheric vortices.  All models account for the radiative balance between heating (methane and H$_2$ absorption) and cooling (from C$_2$H$_2$, C$_2$H$_6$ and, to a much smaller extent, CH$_4$ and the H$_2$ continuum), but they treat the distribution of the stratospheric coolants differently - some using mean abundances that are uniform with latitude and time\cite{14guerlet, 12friedson}; some using meridional distributions that are fixed with time\cite{08greathouse_agu}; and one\cite{16hue} using temporally- and spatially-variable abundances predicted by iterating between their seasonal photochemistry model\cite{15hue} and a radiative model\cite{08greathouse_agu}.  None of these adequately represent the spike in hydrocarbon abundances in the NPSV and SPSV (Fig. \ref{comp_gas}), nor their temporal variation.  To date, only one model\cite{12friedson} includes circulation resulting from seasonal forcings (Supplementary Fig. 2c), making some progress in introducing sharper $\partial{T}/\partial{y}$ gradients, but this model was not designed with polar circulations in mind and does not generate gradients in the right locations.


Warming by polar subsidence of order 1 mm s$^{-1}$ at 1 mbar has been invoked to close the gap between the CIRS measurements and radiative models\cite{15fletcher_poles}.  \added{However, aerosols also constitute a significant part of the radiative budget\cite{14guerlet}, which can introduce a 5-6 K warming effect in the stratosphere\cite{15guerlet}.  Aerosols are a challenge to include in a time-resolved seasonal model,} primarily because the latitudinal distribution and temporal variability of upper tropospheric and stratospheric hazes remain poorly understood.  A recent study of Saturn's reflectivity by Cassini's Imaging Science Subsystem\cite{18requena} observed that the tropospheric hazes moved deeper and became optically thinner poleward of the hexagon. This provides further evidence that subsidence is occurring within the hexagon, and that it extends from the cloud-forming region into the mid-stratosphere.  In addition, the polar region exhibits an increased optical thickness of stratospheric hazes\cite{09west, 18requena} that is potentially related to aerosol production associated with Saturn's aurora.  Saturn's main auroral oval, associated with the boundary between open and closed magnetic field lines, occurs at an average latitude of $76.5^\circ$N\cite{11badman} and therefore completely encompasses the area of the NPSV, sitting $\sim1^\circ$ south of the hexagon latitude ($78^\circ$N).  Stratospheric aerosols, irrespective of how they are formed, may be localised within the NPSV and SPSV and contribute to the sharp rise in stratospheric heating across the vortex boundaries.  Currently it is not possible to disentangle the complementary effects of radiative warming and atmospheric subsidence, \added{although the elevated polar hydrocarbon abundances in Fig. \ref{comp_gas} indicate that polar downwelling occurs within both vortices.}

At nanobar pressures, far above the stratosphere, Saturn's exospheric temperatures increase to $\sim400$ K\cite{15koskinen}, but this value also shows latitudinal and seasonal fluctuations, being warmer in summer (with increased insolation increasing ionospheric conductivity and, therefore, Joule heating rates) than winter.  On Jupiter, stratospheric temperatures are elevated within the auroral oval, both by Joule heating and particulate absorption\cite{18sinclair}.  Future work could address whether the processes heating \added{the atmosphere at} nanobar pressures could also influence the temperatures of Saturn's polar stratosphere at millibar levels, but would require a close coupling of thermospheric and stratospheric circulation, chemistry, and radiative models that include the influences of aerosols.

That the formation and strengthening of the NPSV should be occurring right at the end of Cassini's unprecedented 13-year mission is frustrating.  Nevertheless, ground-based and space-based (James Webb Space Telescope) observations will continue to track its evolving temperature, composition, and aerosol field, as well as the hexagonal boundary, whenever possible.  This northern-summer survey will include searches for any inter-annual variability between the NPSV observations in 1989\cite{89gezari} and those in 2018, one Saturnian year later.



\begin{methods}

\subsection{Low-resolution CIRS time series.}
To track the evolving stratospheric temperature gradients as a function of latitude and time, we adopt a similar technique to previous work\cite{15fletcher_poles, 17fletcher_qpo}, truncating all CIRS interferograms so that they are all treated at the lowest possible spectral resolution of 15 cm$^{-1}$.  This means that all CIRS Saturn observations, including those that rode along with other Cassini instruments, are considered.  These are binned onto a monthly time grid on a $2^\circ$ latitude grid (with a $1^\circ$ step size) for all latitudes, retaining only observations within $\pm10^\circ$ emission angle of the mean for each bin, and ensuring that this emission angle was smaller than $60^\circ$.  Data were also filtered for smearing due to fast motion of the focal plane across the disc; (known interferences; instances of negative radiances introduced by shifting instrument temperatures; and examples where fewer than ten spectra were available at a particular latitude and time. This produced approximately 13,000 spectra for inversion over the 2004-2017 timeline.  \added{At the spatial resolution and sensitivity of the CIRS data, }Saturn's thermal structure is largely axisymmetric in the absence of large-scale storms and waves\cite{15fletcher_poles}, so the coadded spectra are a good approximation of the zonal mean.

\subsection{Spectral Retrieval.}
Spectral inversions use the NEMESIS optimal estimation retrieval algorithm\cite{08irwin}.  The latitude-independent atmospheric prior and weighting between measurement and \textit{a priori} uncertainty are described by Fletcher et al.\cite{15fletcher_poles}, whereas the sources of spectroscopic line data have been updated to those given in Supplementary Table 1 and the H$_2$-H$_2$ collision induced opacities are from the compilation of free-free and dimeric contributions\cite{18fletcher_cia}.  The vertical temperature profile is retrieved alongside a scaling of the \textit{a priori} ethane and acetylene abundances (additional hydrocarbon species are not visible at low spectral resolution, but are considered only at high spectral resolution).  Aerosols are omitted, and we assume the equilibrium distribution of para-hydrogen.  The results for every latitude and month are shown in Fig. \ref{Tpoints}.  Temperature uncertainties range from $\sim3$ K at the tropopause, to 1.5-2.0 K near 200 mbar (the peak of the tropospheric contribution), 3.5-4.5 K at 1-5 mbar (the peak of the stratospheric contribution).  

The resulting time series contains numerous outliers related to problems with calibration, incomplete longitude sampling, differences in spatial resolution, etc\cite{15fletcher_poles}.  Although a quadratic fit to every latitude and altitude was previously used to interpolate and smooth the 10-year time series\cite{15fletcher_poles}, we find that a quadratic was inadequate for the 14-year time series, and instead adopt the tensioned-spline fitting technique\cite{07teanby_splines} as a better representation of the expected sinusoidal variation of temperatures with season.  Splines are best constrained near the middle of the time series and show divergence at the start and end, but are still within the uncertainties of the earliest and latest CIRS measurements available.  

\subsection{Thermal Winds.}
Integration of the thermal wind equation to estimate the zonal winds as a function of latitude and altitude is hampered by several factors\cite{17fletcher_qpo}:  the vertical resolution of nadir spectra is limited to approximately a scale height, implying that regions of strong $\partial{u}/\partial{z}$ are smoothed with altitude;  nadir spectra do not sense $\partial{T}/\partial{y}$ gradients between 5-80 mbar, such that we have \added{no knowledge} of the windshear in the lower stratosphere; and the boundary conditions for the integration (namely, the altitude of the cloud-top winds) remains poorly constrained, and it likely to vary significantly with location on the planet.  By the time the integration reaches millibar pressures, this can lead to uncertainties in the absolute values of the winds of hundreds of metres per second.  \added{Thus we caution the reader that the stratospheric winds derived from nadir remote sounding, in this study and all others, should be considered as only a qualitative guideline for the zonal motions of the middle atmosphere.}

\subsection{Composition from high-resolution CIRS data.}
The spectral resolution of the CIRS data used so far is insufficient for the identification of species beyond CH$_4$, C$_2$H$_2$ and C$_2$H$_6$, so we instead utilise observations acquired at 0.5 cm$^{-1}$ by sitting and staring at a particular latitude (an observation design known as a COMPSIT).  Observation $267\_002$ targeted the NPSV at $85^\circ$N for 8 hours (1200 spectra) with a mean emission angle of $70^\circ$ on March 27th 2017 ($L_s=88.2^\circ$).  This is compared to observation $076\_001$, which targeted the SPSV at $85^\circ$S for 8 hours on July 15th 2008 ($L_s=346^\circ$).  Although the latter was closer to equinox than southern solstice, this was one of the first COMPSITs to target the south polar region, capturing the SPSV in its later stages.  The NPSV spectrum is shown in Fig. \ref{compsit_spx}, with key hydrocarbons labelled.

To reproduce these spectra, $k$-distributions were generated from the gases listed in Supplementary Table 1, and vertical profiles of stratospheric hydrocarbons were extracted from a 2D seasonal photochemistry model\cite{07moses}.  Given that this model includes an estimate for the vertical diffusion ($K_{zz}$) but assumes no meridional motion ($K_{yy}=0$), we elected to take average polar compositions for $72^\circ$N and $72^\circ$S (the maximum poleward extent of the model) over the $L_s=280-90^\circ$ range representing Cassini's mission.  The profiles of methylacetylene (C$_3$H$_4$), diacetylene (C$_4$H$_2$), carbon dioxide (CO$_2$), propane (C$_3$H$_8$) and benzene (C$_6$H$_6$) were simply scaled during the retrievals of full vertical profiles for temperature, ethane and acetylene.  To test the sensitivity of the retrievals to the priors, we set up a grid of 2000 separate inversions for each pole, each with a different profiles for temperature, ethane and acetylene.  The temperature priors merged tropospheric temperatures\cite{09fletcher_ph3} with stratospheric isotherms from 130-170 K.  The acetylene profile was parameterised as a uniform abundance (ranging from 0.5-5.0 ppm) for $1<p<300$ $\mu$bar and a decreasing abundance (linear in log-pressure) into the deeper stratosphere.  The ethane profile was similar, with abundances of 2-20 ppm for $1<p<300$ $\mu$bar and a declining abundance for $p>300$ $ \mu$bar.  These values were broadly representative of those extracted from the photochemical model \cite{07moses}.  

The retrieved profiles, based on this grid of priors, are shown in Fig. \ref{priortest} for both the NPSV ($L_s=88^\circ$) and the SPSV ($L_s=346^\circ$). Some inversions fit the CIRS spectra better than others, but where multiple profiles overlap, we can be confident that it is the data (rather than the prior) that is driving the inversion.  For example, temperatures only begin to diverge for $p<1$ mbar, whereas the gas profiles are best constrained in the 0.5-5.0 mbar range, approximately.  The goodness-of-fit for the acetylene and ethane spectra were calculated from the 710-750 cm$^{-1}$ and 800-840 cm$^{-1}$ ranges respectively, whereas the goodness-of-fit for temperature uses the full spectrum (600-900 cm$^{-1}$  and 1120-1370 cm$^{-1}$).

\end{methods}

\section*{Data availability statement}
All data can be obtained from the primary author (LNF, email: leigh.fletcher@leicester.ac.uk) upon request, or can be accessed from the following GitHub repository \mbox{doi:10.5281/zenodo.1286856}, which contains the temporally and latitudinally averaged spectra used in this study.  Raw and calibrated Cassini Composite Infrared Spectrometer observations are available from NASA's Planetary Data System (PDS).  \added{The entire CIRS database was used in this study, but we provide unique data identifiers where data subsets were used in our figures. The NEMESIS spectral retrieval tool is available upon reasonable request from P.G.J. Irwin (patrick.irwin@physics.ox.ac.uk).  The reconstructed temperature and hydrocarbon fields are also available at the DOI listed above.}


\section*{Supplementary Material}

\begin{figure}
\centering
\makebox[\textwidth][c]{\includegraphics[width=1.3\textwidth]{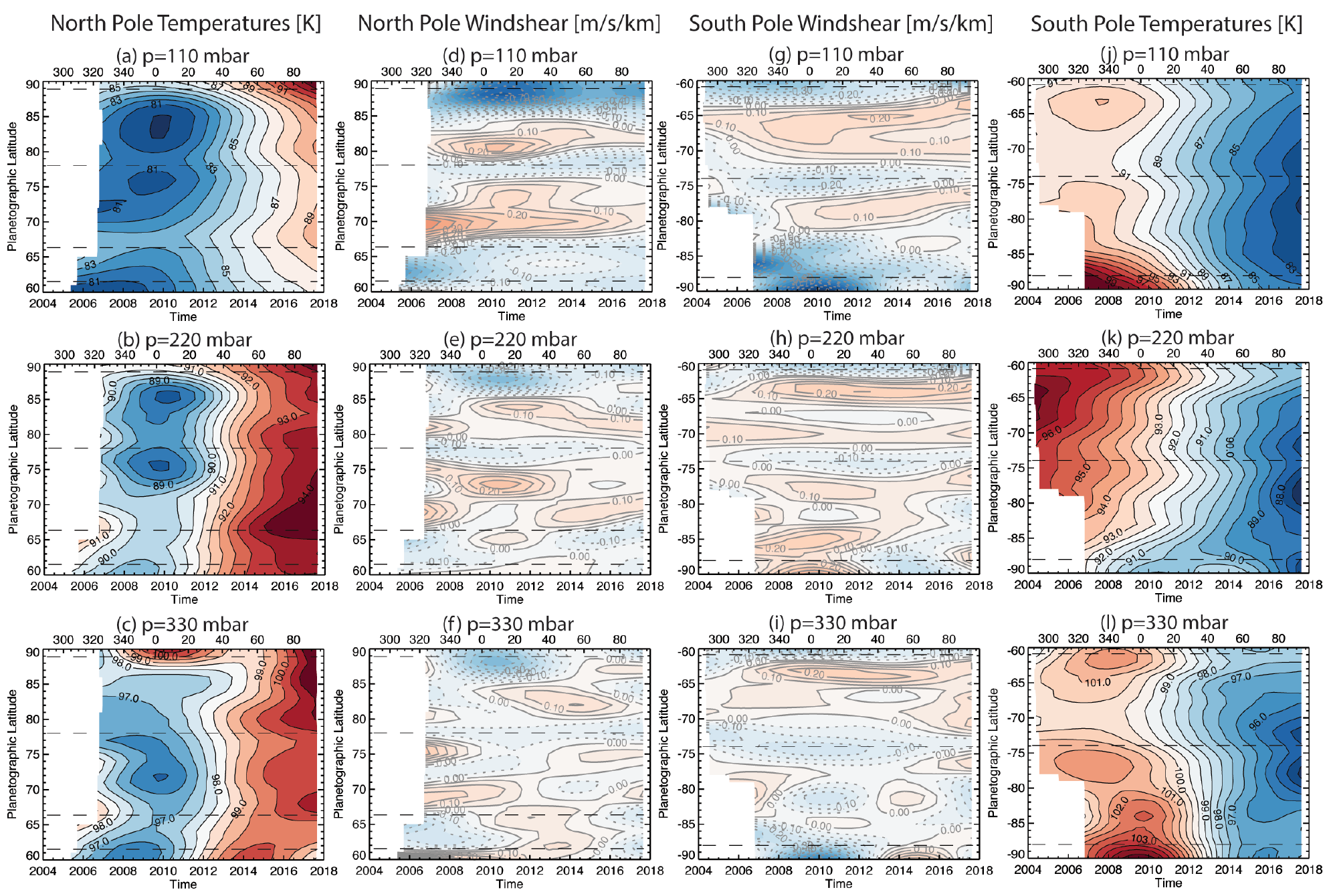}}
\caption{\textbf{Supplementary Fig. 1: North and South polar tropospheric temperature gradients as a function of time throughout the whole Cassini mission, 2004-2017}. We display north polar temperatures (panels a-c), north polar windshears (panels d-f), south polar windshears (panels g-i) and south polar temperatures (panels j-l) at three different pressure levels (110, 220 and 330 mbar).  These were derived from averages of low-resolution CIRS spectra on a monthly temporal grid, and interpolated using tensioned splines\cite{07teanby_splines} to reconstruct a smoothed temperature field.  Horizontal dashed lines signify the peak of eastward zonal jets in the troposphere\cite{15antunano}.  The data are displayed as a function of time (years), but a second horizontal axis provides the planetocentric solar longitude ($L_s$) in degrees.}
\label{tropos_temp}
\end{figure}

\begin{figure}
\centering
\includegraphics[width=7.5cm]{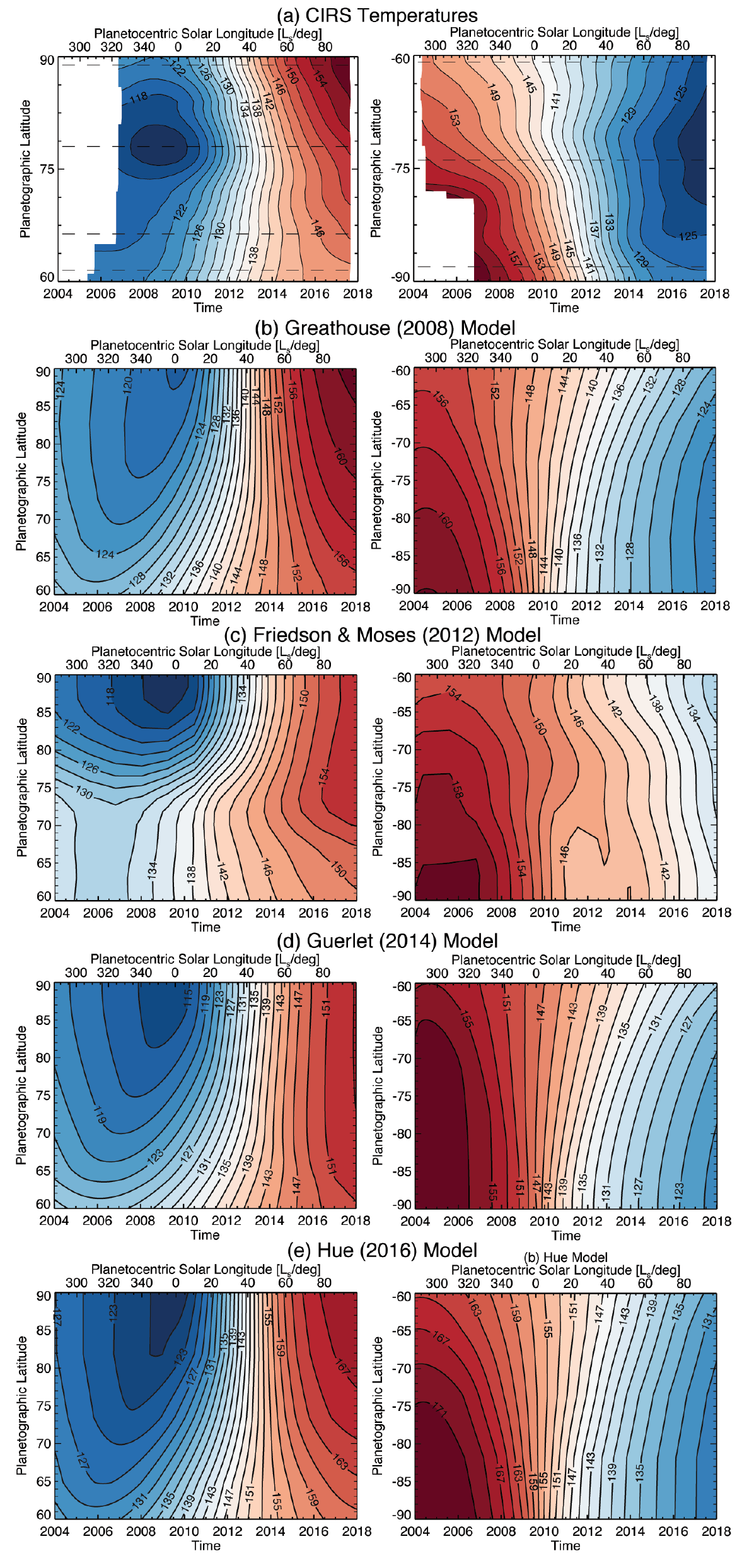}
\caption{\textbf{Supplementary Fig. 2:  Comparison of the 1-mbar temperature field to the predictions of radiative models.}  We compare (a) the CIRS low-resolution measurements to predictions of (b) the radiative model of Greathouse et al.\cite{08greathouse_agu}; (c) the radiative-dynamical model of Friedson and Moses\cite{12friedson}; (d) the radiative-convective model of Guerlet et al.\cite{14guerlet}; and (e) the combined radiative-photochemical model of Hue et al.\cite{16hue}.  Results for the north pole are shown on the left, and the south pole are shown on the right.  Although the predicted temperatures are close to those observed, none of the models include the warming of the North Pole, and none of them predict the appearance of the strong $\partial{T}/\partial{y}$ gradients associated with the polar vortices.}
\label{model_comp}
\end{figure}

\begin{table}
\begin{center}
\caption{Supplementary Table 1:  Sources of spectroscopic linedata and foreign broadening assumptions.  Exponents for temperature dependence $T^n$ given in the final column.}
\begin{tabular}{|p{1cm}| p{3cm}| p{7cm}| p{5cm}| }
\hline
{\bf Gas} & {\bf Line Intensities} & {\bf Broadening Half Width} & {\bf Temperature Dependence} \\
\hline
CH$_4$, CH$_3$D & Brown et al.\cite{03brown} & H$_2$ broadened using a half-width of 0.059 cm$^{-1}$atm$^{-1}$ at 296 K & $n=0.44$ Margolis et al.\cite{93margolis} \\
\hline
C$_2$H$_6$& Vander-Auwera et al.\cite{07vander} & 0.11 cm$^{-1}$atm$^{-1}$ at 296 K Blass et al.\cite{87blass} &  $n=0.94$ Halsey et al.\cite{88halsey} \\
\hline
C$_2$H$_2$ & GEISA'03\cite{05geisa} & Fits to data in Varanasi et al.\cite{92varanasi} & Varanasi et al.\cite{92varanasi} \\
\hline
PH$_3$ & Kleiner et al.\cite{03kleiner} & Broadened by both H$_2$ and He using $\gamma_{H_2}=0.1078-0.0014J$ cm$^{-1}$atm$^{-1}$ and $\gamma_{He}=0.0618-0.0012J$ cm$^{-1}$atm$^{-1}$ Levy et al.\cite{93levy}, Bouanich et al.\cite{04bouanich} &   $n=0.702-0.01J$  ($J$ is the rotational quantum number) Salem et al.\cite{04salem} \\
\hline
NH$_3$ & Kleiner et al.\cite{03kleiner} & Empirical model of Brown et al.\cite{94brown} & Brown et al.\cite{94brown} \\
\hline
C$_2$H$_4$ & GEISA'03\cite{05geisa} & Polynomial fits to data from Bouanich et al.\cite{03bouanich_c2h4, 04bouanich_c2h4} & $n=0.73$, Bouanich et al.\cite{04bouanich_c2h4} \\
\hline
C$_3$H$_4$ & GEISA'09\cite{11jacquinet} & 0.075 cm$^{-1}$atm$^{-1}$ for all lines & $n=0.50$ assumed \\
\hline
C$_4$H$_2$ & GEISA'09\cite{11jacquinet} & 0.1 cm$^{-1}$atm$^{-1}$ for all lines & $n=0.75$ assumed \\
\hline
C$_3$H$_8$ & GEISA'09\cite{11jacquinet} & 0.08 cm$^{-1}$atm$^{-1}$ for all lines & $n=0.75$ assumed. \\
\hline
C$_6$H$_6$ &GEISA'09\cite{11jacquinet} & Linear fit to N$_2$-broadening of Waschull et al.\cite{98waschull} & $n=0.75$ assumed \\
\hline
\end{tabular}
\end{center}

\label{tab:linedata}
\end{table}



\pagebreak

\section*{References}
\bibliography{references}


\begin{addendum}
 \item [Correspondence] Requests for materials should be addressed to L.N.F. (leigh.fletcher@le.ac.uk).
 \item [Acknowledgements]  This work is based on data acquired by the Cassini Composite Infrared Spectrometer, and would not have been possible without the tireless efforts of the instrument design, operations, and calibration team over more than two decades.  LNF was supported by a Royal Society Research Fellowship and European Research Council Consolidator Grant (under the European Union's Horizon 2020 research and innovation programme, grant agreement No 723890) at the University of Leicester.  The UK authors acknowledge the support of the Science and Technology Facilities Council (STFC).  A portion of this work was performed by GSO at the Jet Propulsion Laboratory, California Institute of Technology, under a contract with NASA.  JAS was supported by the NASA Postdoctoral and Caltech programs as well as by a contract with NASA.  SG was supported by the Centre National d'Etudes Spatiales (CNES). This research used the ALICE High Performance Computing Facility at the University of Leicester.  We are extremely grateful to S. Guerlet, V. Hue. A.J. Friedson, T. Greathouse and J.I. Moses for sharing the outputs of their Saturn simulations.
 \item[Competing Interests] The authors declare that they have no competing interests.
 \item[Author Contributions] LNF wrote the manuscript and performed the analysis.  GSO, JAS, SG, AA participated in the data analysis; PLR, RKA, FMF aided in dynamical interpretations; PGJI and SBC developed the software for spectral modelling; GLB, JH, BEH, MS designed and implemented the CIRS observations; NG and AM developed the calibration software.  All of the authors discussed the results and commented on the manuscript.
\end{addendum}





\end{document}